\documentclass[final,onecolumn,prf,aps,amsfonts,amssymb,showpacs,superscriptaddress,floatfix,nofootinbib]{revtex4-1}
\usepackage{amsmath}
\usepackage{graphicx}
\usepackage[dvipsnames]{color}
\usepackage{hyperref}
\usepackage{psfrag}
\usepackage{stmaryrd}
\usepackage{amssymb}
\usepackage{wasysym}
\usepackage{float}
\usepackage{color}

\usepackage[latin1]{inputenc}
\usepackage[T1]{fontenc}
\usepackage[french,english]{babel}
\usepackage{lmodern}

\usepackage{geometry}
\begin{document}

\title{Linear and nonlinear regimes of an inertial wave attractor}

\author{Maxime Brunet}
\affiliation{Laboratoire FAST, CNRS, Universit\'e Paris-Sud,
Universit\'e Paris-Saclay, 91405 Orsay, France}
\author{Thierry Dauxois}
\affiliation{Université de Lyon, ENS de Lyon, Université Claude Bernard, CNRS,
    Laboratoire de Physique, F-69342 Lyon, France}
\author{Pierre-Philippe~Cortet}
\email[]{ppcortet@fast.u-psud.fr} \affiliation{Laboratoire FAST,
CNRS, Universit\'e Paris-Sud, Universit\'e Paris-Saclay, 91405
Orsay, France}

\date{\today}

\begin{abstract}
We present an experimental analysis of the linear and non-linear
regimes of an attractor of inertial waves in a trapezoidal cavity
under rotation. Varying the rotation rate and the forcing
amplitude and wavelength, we identify the scaling laws followed by
the attractor amplitude and wavelength in both regimes. In
particular, we show that the non-linear scaling laws can be well
described by replacing the fluid viscosity in the linear model by a turbulent
viscosity, a result that could help
extrapolating attractor theory to geo/astrophysically relevant
situations. We further study the triadic resonance instability of
the attractor which is at the origin of the turbulent viscosity. We
show that the typical frequencies of the subharmonic waves
produced by the instability behaves very differently from
previously reported numerical results and from the prediction of
the theory of triadic resonance. This behavior might
be related to the deviation from horizontal invariance of the
attractor in our experiment in relation with the presence of
vertical walls of the cavity, an effect that should be at play in
all practical situations.
\end{abstract}

\maketitle

\section{Introduction}

Fluids submitted to a global rotation enable the propagation of a
specific class of waves, called inertial
waves~\cite{Greenspan1968} as a result of the restoring action of
the Coriolis force. In an inertial wave, energy propagates in a
direction tilted by an angle $\theta$ with respect to the
horizontal which is defined by the dispersion relation
\begin{eqnarray}
\sigma=2\Omega\cos\theta,
\end{eqnarray}
where $\sigma$ is the wave angular frequency and $\Omega$ the
fluid rotation rate around the vertical axis. Inertial waves are
cousins of internal waves of gravity propagating in linearly
stratified fluids~\cite{Pedlosky1987}: they have similar
dispersion relations linking the ratio between the wave frequency
and the rotation rate or the buoyancy frequency to a specific
direction along which their energy propagates. These dispersion
relations lead to orthogonal group and phase velocities and also
let the lengthscales of the wave unprescribed by the frequency.
These lengthscales (wavelength, beam width) are consequently set
by boundary conditions, viscous dissipation and eventually
non-linearities. This leads to a variety of wave structures like
self-similar wave
beams~\cite{Mowbray1967,Flynn2003,Cortet2010,Machicoane2015},
plane waves~\cite{Mercier2010,Bordes2012} or resonant cavity
modes~\cite{Aldridge1969,McEwan1970,Maas2003b,Boisson2012,Boisson2012b}.
These waves are relevant in geophysics and astrophysics in which
they often merge into inertia-gravity waves with a single
dispersion relation coupling rotation and
buoyancy~\cite{Lighthill1978,Pedlosky1987}. Considering pure
inertial and pure gravity waves, a major difference however
exists: gravity waves involve rectilinear fluid oscillations in
the plane tilted by the angle $\theta$ whereas in inertial waves
fluid particles describe anti-cyclonic circular translations in
this plane.

In closed domains, modes of standing waves can be found for
specific frequencies when the walls are normal and parallel to the
rotation axis~\cite{Greenspan1968,McEwan1970} but also in some
geo/astrophysically relevant geometries such as spheres and
spheroids~\cite{Zhang2004}. In geo and astrophysics, several types
of global forcing may be at the origin of such modes. For example,
modes can be excited in spheres and spherical shells by a
longitudinal
libration~\cite{Aldridge1969,Rieutord1991,Tilgner1999,Noir2009,Calkins2010}
consisting in a time modulation of the rotation rate.
Precession~\cite{Busse1968,Kerswell1995,Noir2001,Boisson2012} and
tidal deformation of the planet crust~\cite{Suess1971,Morize2010}
are other examples.

However, closed domains generally include sloping walls in which
case the wave focusing and defocusing induced by the peculiar
reflection laws of inertial/internal gravity
waves~\cite{Phillips1963} prevent the existence of cavity
eigenmodes~\cite{Maas1995}. It is a consequence of the dispersion
relation: the wave keeps constant its propagation angle (in
absolute value) with respect to the horizontal when reflecting on
a wall. This implies for reflection on tilted walls that
Snell-Descartes laws are not verified and that the wave
lengthscales are enhanced or reduced depending on the fact the
wave is descending or climbing the slope~\cite{Phillips1963}. This
peculiar physics can lead to the emergence of limit cycles, called
wave attractors~\cite{Maas1995,Maas1997,Rieutord2001,Manders2003},
on which the waves concentrate when excited in closed domains with
tilted walls, including the astrophysically relevant case of
spherical shells~\cite{Rieutord2001}. From a theoretical point of
view, tracing rays respecting the reflection laws directly reveals
the existence (or the absence) of an attractor for a given
wave frequency, via the convergence (or not) of all rays towards a
unique limit cycle. Inviscid attractors actually exist over
specific frequency ranges depending on the cavity
geometry~\cite{Maas1995,Maas1997,Rieutord2001,Manders2003}.

The first clear experimental observations of attractors were done
in trapezoidal cavities by Maas and co-workers both for internal gravity
waves~\cite{Maas1997} and for inertial
waves~\cite{Maas2001,Manders2003,Manders2004} (and also latter in another
geometry by Klein
\textit{et al.}~\cite{Klein2014} for inertial waves). In these experiments, the
wave forcing was realized through a global motion of the water tank: a
longitudinal libration in the case of rotation and an oscillating
vertical translation in the case of stratification. In reality,
the energy injected at the forcing frequency $\sigma_0$ does not
focus on an infinitely thin parallelogram because of viscous
dissipation. The scales (width and wavelength) of the attractor then follow
from the
competition of the energy focusing during reflections on tilted
walls and the energy dissipation during wave propagation: focusing
reduces scales whereas dissipation preferentially damps small
scales. It has been proposed that the attractor wave beam once
unwrapped can be described as a self-similar wave beam emitted by
a virtual point source located upstream of the focusing
reflection~\cite{Rieutord2001,Ogilvie2005,Hazewinkel2008,Grisouard2008}.
This model, in which the scale reduction at the focusing
reflection is exactly compensated by the viscous spreading of the
beam emitted by the point source~\cite{Cortet2010,Machicoane2015},
has been tested with an increasing success in numerical
simulations~\cite{Grisouard2008,Jouve2014}.

Since in geo/astrophysics extremely large Reynolds numbers are
involved, a particularly interesting, but not thoroughly
discussed, facet of this problem is how the attractor is affected
by non-linearities when increasing the forcing amplitude. This
question has started to be explored experimentally by
Scolan~\textit{et al.}~\cite{Scolan2013} and Brouzet~\textit{et
al.}~\cite{Brouzet2017} in stratified fluids and by Jouve and
Ogilvie~\cite{Jouve2014} in direct numerical simulations of a
rotating tilted square: they have revealed the emergence of an
instability of the attractor feeding two subharmonic
waves in triadic resonance with the attractor. This non-linear
process has further been shown to affect the attractor by damping
its amplitude and increasing its scales. These observations have
been proposed to result from the additional effective dissipation
that the instability induces for the
attractor~\cite{Jouve2014,Brouzet2017}. However, no scaling laws
have up to now been identified to describe the non-linear
evolution of the attractor. Even in the linear regime, it is not
clear how the attractor amplitude scales with the forcing Reynolds
and Rossby numbers.

In this article, we propose such scaling laws and compare them
successfully with experimental data. We explore the linear and
non-linear regimes of an inertial wave attractor in a trapezoidal
cavity under rotation using a co-rotating particle image
velocimetry system. The flow is generated by a deformable upper
cover of the cavity which has a sinusoidal shape with a phase
propagating horizontally. The dependencies of the scale and
amplitude of the attractor are studied as a function of the
forcing amplitude for several rotation rates and forcing
wavelengths. This large data set allows us to validate scaling
laws for the attractor in the non-linear regime which have been
theoretically established assuming that the linear attractor model
is still valid in the non-linear regime if one replaces the fluid
viscosity by a turbulent viscosity. We further report a
detailed study of the subharmonic waves produced by the triadic
resonance instability responsible for the attractor non-linear
evolution mentioned above. We show that, as the forcing Reynolds
number is increased, the two typical frequencies, at which
subharmonic waves are produced, are moving away from half the
attractor frequency, in contradiction with what is expected from
the theory of the triadic resonance instability as well as from numerical
simulations of inertial wave attractors~\cite{Jouve2014}. We
suggest that these differences could be related to the
three-dimensionality of fluid motions in inertial waves interacting
here with the cavity vertical walls. We finally show that a
significant amount of the waves produced by the instability are
three-dimensional, breaking the invariance in one horizontal direction of the
forcing and of the attractor, which could also be a consequence of
the waves interaction with the vertical walls.

\section{Experimental setup}

\begin{figure}
    \centerline{\includegraphics[width=10.5cm]{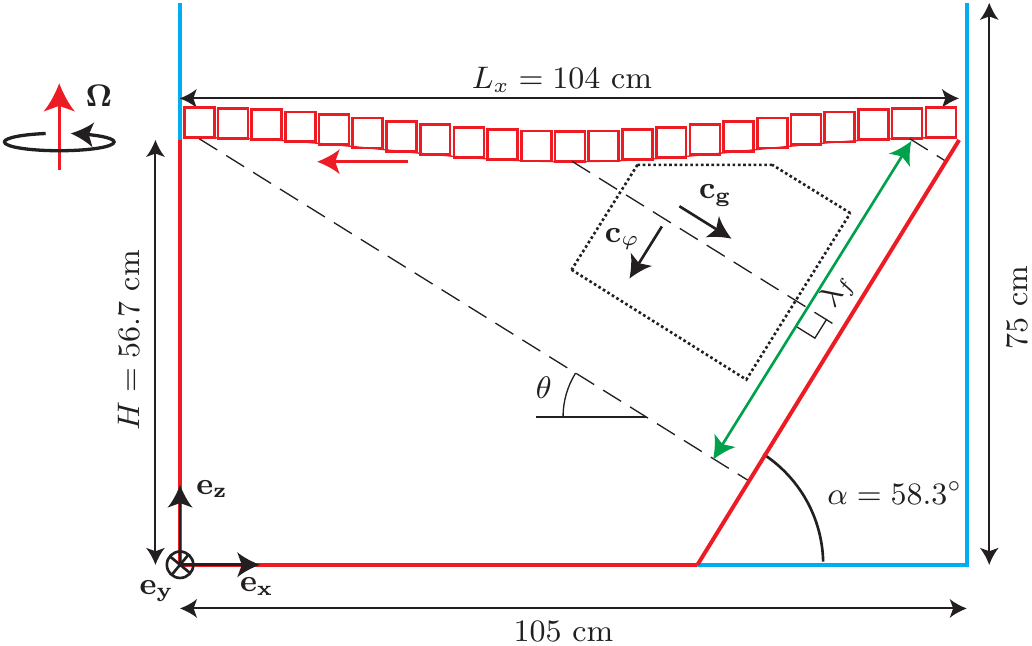}}
    \caption{Scheme of the water tank (in blue) containing the
    trapezoidal cavity (in red). The upper wall of the cavity is made
    of a series of $n = 23$ horizontal bars, each of them connected to
    a linear motor able to drive it in a vertical motion. The
    truncated dashed rectangle shows the region over which the
    velocity amplitude reported in Fig.~\ref{fig:forcing} is
    averaged.}\label{fig:cavite}
\end{figure}

The flow is generated in a trapezoidal cavity of height
$H=56.7$~cm, length $L_x=104$~cm and width $L_y=105$~cm as
illustrated in Fig.~\ref{fig:cavite}. This cavity is contained in
a parallelepipedic tank of $105\times 105$~cm$^2$ base and $75$~cm
height filled with $63$~cm of water. One wall of the cavity is a
plate tilted by an angle $\alpha=58.3^\circ$ with respect to the
horizontal. The forcing of the flow is realized by the upper wall
of the cavity which is made of a series of $23$ horizontal bars,
$86$~cm long and centered in the tank in the $y$ direction. The
bars have a square section of $40$~mm side in the $(x,z)$ plane
and are spaced of $5$~mm in the $x$ direction. Each of these bars
is connected to a linear motor able to drive it in a vertical
motion. This wavemaker imposes the upper cover of the flow to
approximate the following wavy shape
\begin{eqnarray} \label{eq:upperwall}
Z(x,t)=H+A\left[\cos\left(\sigma_0 t + k_0 x\right)-1\right],
\end{eqnarray}
where $k_0=2\pi/\lambda_0$ and two values of the wavelength have
been considered: $\lambda_0=11 \times 4.5$~cm$=49.5$~cm and
$\lambda_0=22 \times 4.5$~cm$=99$~cm ($4.5$~cm is the size of an oscillating
bar plus the interval between two bars). The whole system is mounted on a 2\,m
diameter platform
rotating at a constant rate $\Omega=3$~rpm or $18$~rpm about the
vertical axis~$z$. The rotation of the platform is set at least 30
minutes before the wavemaker is started to avoid transient spin-up
recirculations. The angular frequency of the wavemaker is set to
$\sigma_0=0.85 \times 2\Omega$. The amplitude $A$ of the bars
motion $Z(x,t)$ is varied in the range [$0.09$~mm,~$18$~mm].

\begin{figure}
    \centerline{\includegraphics[width=9.5cm]{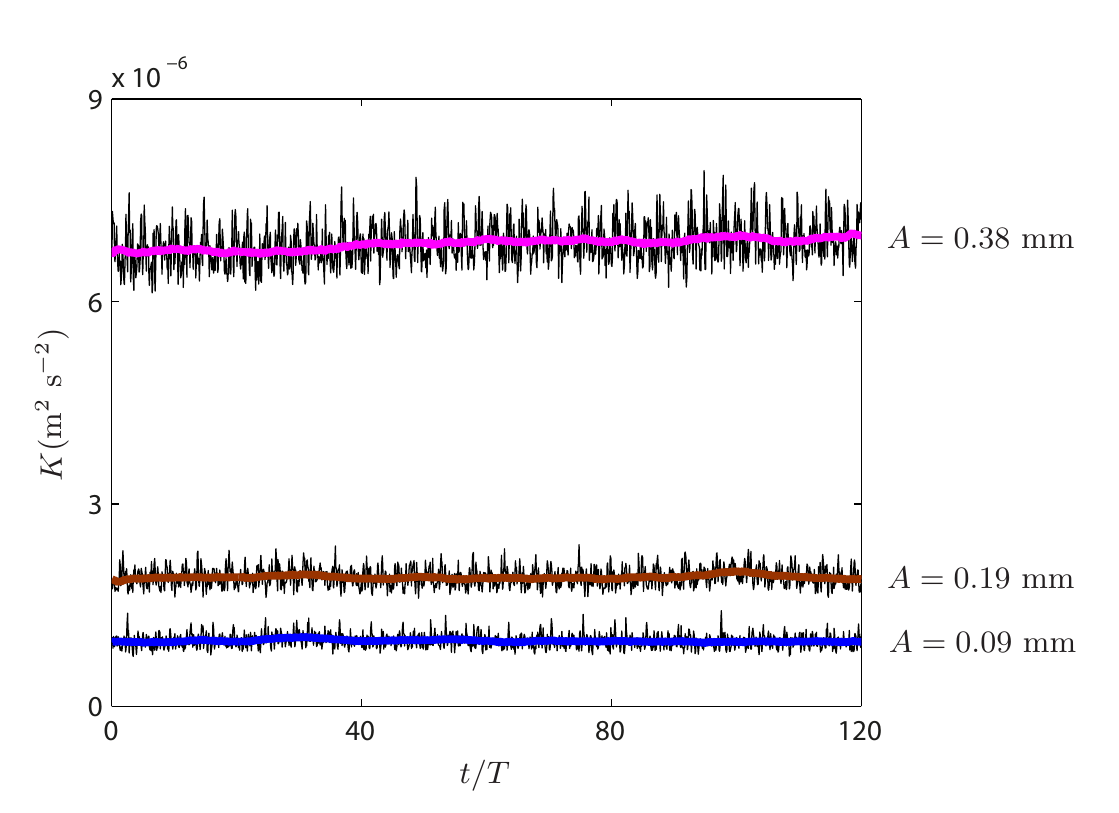}}
    \caption{Time series of the spatially averaged kinetic energy 
    $K=\langle u_x^2+u_z^2 \rangle/2$ for
    $\Omega=18$~rpm, $\lambda_f=52.4$~cm and three forcing amplitudes
    $A=0.09$~mm, $A=0.19$~mm and $A=0.38$~mm. Time $t$ is normalized by
    the period $T=2\pi/\sigma_0$ of the wavemaker. For each
    time series the black thin line shows the original data and the
    thick line its moving average using a time window of $5T$.}\label{fig:stat}
\end{figure}

The two components $(u_x,u_z)$ of the velocity field are measured
in the vertical plane $y=y_0 = L_y/3$ using a particle image
velocimetry (PIV) system mounted in the rotating frame ($y=0$ is
the front side of the tank). The fluid is seeded with 10~$\mu$m
tracer particles and illuminated by a laser sheet generated by a
corotating 140 mJ Nd:YAG pulsed laser. Pairs of images of
particles are acquired using two $2\,360 \times 1\,776$~pixels
cameras. Using a spatial calibration, the two images in each pair
are combined into a single image covering the whole trapezoidal
cavity. For wavemaker amplitudes $A\leq 1.50$~mm, image
acquisition consists of series of $1\,440$ to $5\,760$ image pairs
recorded at a rate between $1.5$ and $24.4$~Hz depending on the
wavemaker amplitude $A$ and on the rotation rate $\Omega=3$~rpm or
$18$~rpm. These values correspond to the acquisition of
$120$~periods of the wavemaker with a time resolution between $12$
and $48$~image pairs per wavemaker period. For the wavemaker
amplitudes $A$ larger than $1.5$~mm, acquisitions consist in the
regular recording of two pairs of images separated by a time
interval $dt\in[9$~ms$,~29$~ms$]$. This double-frame PIV
configuration is rendered necessary by the large amplitude of the
fluid velocity. For these large values of $A$, $120$ to
$360$~periods of the wavemaker are recorded with a time resolution
of $12$ doublets of image pairs per wavemaker period. We finally
compute cross-correlation between successive images over windows
of $32 \times 32$~pixels with $50\%$ overlap. This produces
velocity fields of spatial resolution $4.17$~mm, with $130$ lines
of between $164$ (at the bottom) and $244$ (at the top) vectors
almost covering the whole section of the cavity. A few
acquisitions have been realized during the transient settling of
the flow after the wavemaker is started. They have revealed that a
steady state is reached after a few dozen periods of the wavemaker
forcing. In the data discussed in the following, image acquisition
is started after 500~periods of the wavemaker oscillation. In
order to illustrate the degree of statistical stationarity of the
flow, we report in Fig.~\ref{fig:stat}, the time series of the
kinetic energy $K=\langle u_x^2+u_z^2 \rangle/2$ for three
experiments, the angular brackets denoting the spatial average
over the measurement plane.

In an unbounded domain, our wavemaker is expected to generate an
inertial wave with its energy propagating at an angle
$\theta=\cos^{-1}(\sigma_0/2\Omega)\simeq 32.0^\circ$ with respect
to the horizontal and of wavelength $\lambda_f=\lambda_0
\sin\theta=26.2$~cm or $52.4$~cm (see Fig.~\ref{fig:cavite}). In
line with Eq.~(\ref{eq:upperwall}), the phase of the wavemaker
propagates toward decreasing values of $x$, selecting the
excitation of a wave with an energy propagating toward increasing
$x$ (see the group $\bf c_g$ and phase $\bf c_\varphi$ velocities
in Fig.~\ref{fig:cavite}). A more subtle point is to predict the
velocity amplitude of the raw wave excited by the wavemaker.
Considering the small thickness of the Ekman boundary layers
expected on the wavemaker, $\delta=\sqrt{\nu/2\Omega} \simeq
0.5$~mm, one can assume that an effective free-slip condition
holds at the scale of the wave~\cite{Machicoane2018}. One can
therefore expect that the wavemaker prescribes the $z$-component
of the velocity field. Since in an inertial wave fluid particles
describe anticyclonic circular translation in planes tilted by the
angle $\theta$~\cite{Greenspan1968,Bordes2012,Machicoane2018}, the
amplitude of the wave velocity components (along directions
$\theta$ and $y$) are expected to be $U_f =
A\sigma_0/\sin\theta\in [0.6,~36.2]$~mm~s$^{-1}$. This prediction
is expected to hold when $\theta \rightarrow 90^\circ$ but probably
fails when $\theta \rightarrow 0^\circ$ since it would imply a
divergence of the wave velocity amplitude.

\begin{figure}
    \centerline{\includegraphics[width=7cm]{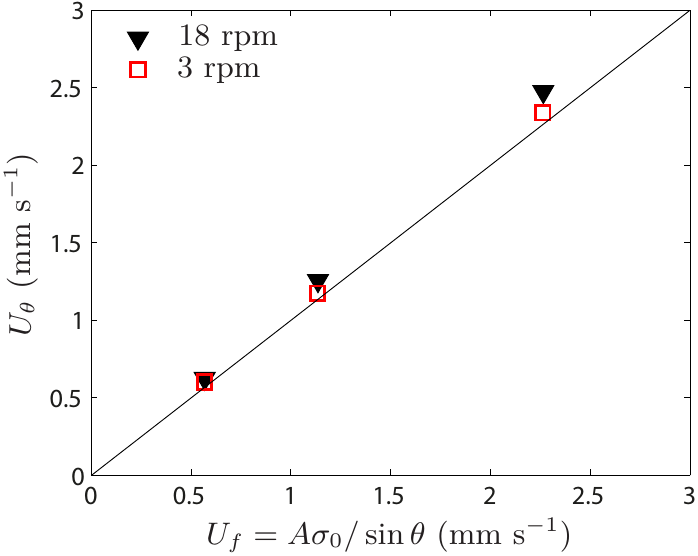}}
    \caption{Velocity amplitude $U_\theta$ of the wave excited by the
    generator in the cavity in absence of the tilted plane for
    $\Omega=3$~rpm and $18$~rpm, $\lambda_f=52.4$~cm and for three
    forcing amplitudes $A=0.09$, $0.19$ and $0.36$~mm. The data is
    reported as a function of $U_f=A\sigma_0/\sin\theta$. The solid
    line indicates the identity function.}\label{fig:forcing}
\end{figure}

To confirm that this prediction $U_f = A\sigma_0/\sin\theta$ holds
at the angle $\theta \simeq 32.0^\circ$ used in the experiments,
we have realized a few complementary experiments without the
tilted plane. In Fig.~\ref{fig:forcing}, we report the velocity
oscillations amplitude $U_\theta$ along the direction $\theta$,
i.e. the direction of the wave group velocity, as a function of
the predicted amplitude $U_f= A\sigma_0/\sin\theta$. To get
$U_\theta$, we first Fourier filter the velocity field obtained
from particle image velocimetry at frequency $\sigma_0$. We then
compute the amplitude of the velocity oscillations along direction
$\theta$ and finally take the spatial average of the resulting
field over the region delineated by the dashed lines in
Fig.~\ref{fig:cavite} in the center of the excited wave.
Figure~\ref{fig:forcing} confirms that the excited wave amplitude
is indeed close to $U_f=A\sigma_0/\sin\theta$ within a $10\%$
precision which confirms the relevance of our estimate. The values
found for $U_\theta$ are actually slightly larger than $U_f$.
This discrepancy can be the consequence of the fact that the
assumptions made to model $U_f$, i.e. the free-slip boundary
condition and the fact that the wavemaker is watertight (which is
not the case because of the 5~mm gap between the bars), are only
valid to the first order. This discrepancy might also reveal the
contribution to the measured velocity amplitude $U_\theta$ of the
wave after its reflections on the four cavity walls which is
superimposing to the original wave. In the following, $U_f$ and
$\lambda_f$ will be named forcing velocity and forcing wavelength
and $Re_f=U_f\lambda_f/\nu=A\sigma_0\lambda_0/\nu$ the forcing
Reynolds number.

\section{Wave attractor in the linear regime and beyond}\label{sec:theo}

\begin{figure}
    \centerline{\includegraphics[width=5cm]{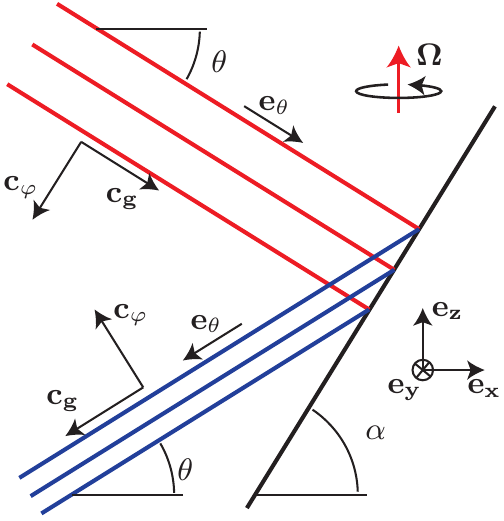}}
    \caption{Reflection of an inertial wave of angular frequency
    $\sigma=2\Omega \cos\theta$ on a plane tilted by an angle
    $\alpha$. The vectors $\bf c_\varphi$ and $\bf c_g$ show the directions of
    the phase and group velocities for the
    incident and reflected beams.}\label{fig:reflect}
\end{figure}

Attractors of internal waves of inertia or gravity can emerge in
cavities with walls non-parallel or non-normal to the rotation or
stratification axis. This is due to the anomalous reflection laws
of these waves which keep constant their propagation
angle~$\theta$ with respect to the horizontal during a
reflection~\cite{Phillips1963}. As a consequence, when reflecting
downwards on a wall tilted by an angle $\alpha$ (see
Fig.~\ref{fig:reflect}), a wave has its transverse lengthscales
reduced by a factor
\begin{eqnarray}\label{eq:gamma}
\gamma=\frac{\sin(\alpha+\theta)}{\sin(\alpha-\theta)}(=2.25~\textrm{here
in the experiments}).
\end{eqnarray}
Moreover, noting that the group velocity of a plane inertial wave
of angular frequency $\sigma= 2\Omega \cos\theta$ has a magnitude
$|{\bf c_g}|=\Omega \lambda \sin\theta/\pi$, the conservation of
the energy flux implies that $U_{\theta} \lambda$ and $U_{y}
\lambda$ are conserved at reflection (at least in the inviscid
case~\cite{Beckebanze2018}), $\lambda$ being the wavelength and
$U_\theta$ and $U_y$ the amplitude of the velocity oscillations
along directions ${\bf e_\theta}$ and ${\bf e_y}$ respectively
(see Fig.~\ref{fig:reflect}). Besides, the wave vorticity scaling
as $U_{y,\theta}/\lambda$ is amplified by a factor $\gamma^2$ at
the reflection.

In a closed domain, this focusing process leads, for certain
geometry-dependent ranges of angles $\theta$, to an energy
concentration on a wave
attractor~\cite{Maas1995,Maas1997,Rieutord2001,Manders2003}. In
the trapezoidal cavity considered here, a $(1,1)$~inertial wave
attractor is expected when waves are generated with a propagation
angle $\theta$ in the range between
\begin{eqnarray}
\theta_1&=&\tan^{-1}(H/L_x)\simeq 28.6^\circ, {\rm and}\\
\theta_2&=&\tan^{-1}\left(\frac{H\tan(\alpha)}{L_x\tan(\alpha)-H}\right)\simeq
39.4^\circ.
\end{eqnarray}
The notation $(1,1)$ corresponds to the simplest class of
attractor in a trapezoidal cavity and follows from the
nomenclature introduced in~\cite{Maas1997}, the first number being
the number of reflections on the bottom wall and the second the
number of reflections on the sloping wall. The two limit angles
$\theta_1$ and $\theta_2$ correspond to the slope of the diagonals
of the trapezoidal cavity.

\begin{figure}
    \centerline{\includegraphics[width=10.5cm]{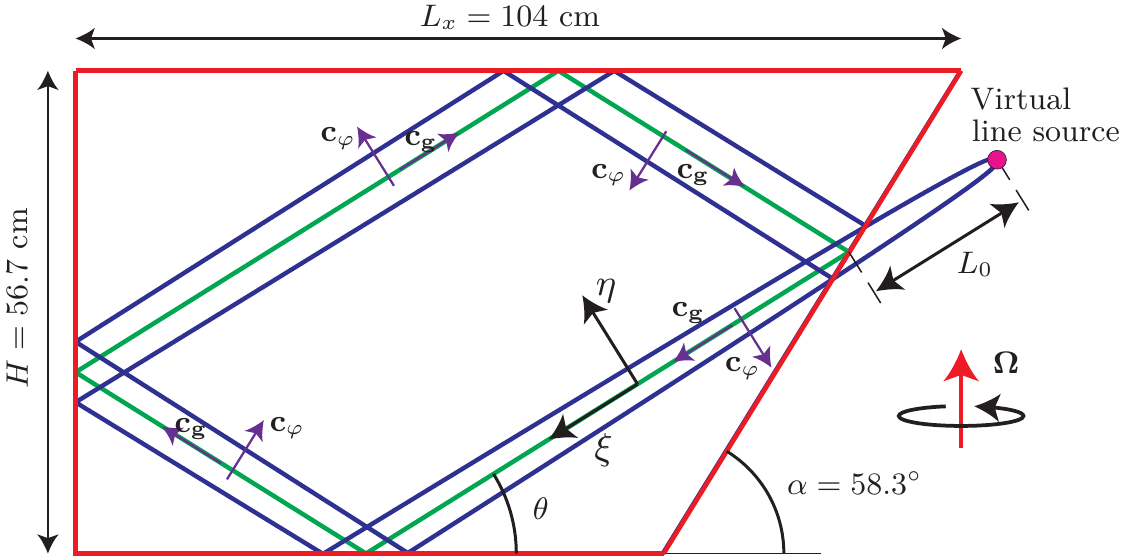}}
    \caption{Scheme of the theoretical inviscid (in green) and viscous
    (in blue) attractors in the trapezoidal cavity, for
    $\theta=32.0^\circ$. The blue lines more precisely show the width
    at mid-height of the velocity envelope of the
    Moore-and-Saffman self-similar wave beam
    (Eq.~\ref{eq:mooresaff}). The vectors $\bf c_\varphi$ and $\bf c_g$ show
    for each attractor branch the direction of the phase and group
    velocities.}\label{fig:cavite_attract}
\end{figure}

In Fig.~\ref{fig:cavite_attract}, we show for the angle
$\theta\simeq 32.0^\circ$ used in the experiments the unique
closed parallelogram with its vertices on the walls of the cavity
and its sides all tilted by an angle~$\theta$. This parallelogram
corresponds to the inviscid wave attractor: ray tracing for any
wave at $\sigma_0=2\Omega\cos\theta$ will converge toward this
parallelogram. The rate of convergence of the rays toward the
inviscid attractor in a trapezoidal cavity has been characterized
by Maas and co-workers in~\cite{Maas1997} via the calculation of
the Lyapunov exponents for each couple of non-dimensional
geometric parameters ($d=1-2H/(L_x\tan
\alpha),\tau=2H/(L_x\tan\theta)$). In~\cite{Maas1997}, internal
gravity waves are considered, but since inertial waves behave
exactly the same way regarding two-dimensional ray tracing, the Lyapunov
exponents diagram as a function of ($d,\tau$) is expected to be identical in
our case. This diagram reveals regions of strong convergence
called Arnold tongue~\cite{Maas1997,Brouzet2017b}. In our
experiments, one has ($d \simeq 0.33,\tau\simeq1.74$) which falls
in the middle of the Arnold tongue corresponding to the
$(1,1)$-attractor.

In a viscous fluid, the inviscid concentration of energy on a line
attractor is prevented by viscous dissipation of waves during
their propagation. The structure of the attractor can eventually
be seen as the result of a balance between focusing at the sloping
wall and the viscous spreading of a self-similar polychromatic
wave
beam~\cite{Rieutord2001,Ogilvie2005,Hazewinkel2008,Grisouard2008}.
As proposed in~\cite{Grisouard2008} for internal wave attractors
in a linearly stratified fluid, the four branches of the
attractor, once unwrapped, can indeed be seen as part of a beam
emitted by a virtual line source (invariant along the
$y$~direction), located at a distance $L_0$ upstream of the
focusing reflection (see Fig.~\ref{fig:cavite_attract}).

From a general point of view, the wave beam excited by a line
source has a self-similar transverse structure. Its
velocity component along the propagation direction $\theta$ is
given by (see Ref.~\cite{Machicoane2015} for details)
\begin{eqnarray}\label{eq:mooresaff}
u_\xi^{(m)}(\xi,\eta,t)=U_0
\left(\frac{\ell}{\xi}\right)^{(m+1)/3}\left[c_m(\eta/\delta)\cos(\sigma
t+\phi)+s_m(\eta/\delta)\sin(\sigma t+\phi)\right],
\end{eqnarray}
and the out-of-plane vorticity component by
\begin{equation}
\omega_y^{(m)}(\xi,\eta,t) = \frac{U_0}{\ell} \left(
\frac{\ell}{\xi} \right)^{(m+2)/3} \left[-s_{m+1}(\eta/\delta)
\cos (\sigma t +\phi) + c_{m+1}(\eta/\delta) \sin (\sigma t +
\phi)\right], \label{eq:wy}
\end{equation}
where $\xi$ is the distance from the source, $\eta$ the local
transverse coordinate,
\begin{eqnarray}\label{eq:ell}
\ell=\frac{\nu^{1/2}}{((2\Omega)^2-\sigma_0^2)^{1/4}}
\end{eqnarray}
a viscous scale and
\begin{eqnarray}\label{eq:delta}
\delta(\xi)=\xi^{1/3}\ell^{2/3}
\end{eqnarray}
the scaling law followed by the width (and all other transverse
length scales) of the beam. In the experiments, $\ell\simeq
1.74$~mm for $\Omega=3$~rpm and $0.71$~mm for $\Omega=18$~rpm. The
functions $c_m$ and $s_m$ have been introduced by
Moore and Saffman~\cite{Moore1969} and Thomas and
Stevenson~\cite{Thomas1972} to describe self-similar wave beams of
inertial and internal waves respectively. These real functions are
defined by
\begin{eqnarray}\label{eq:msf}
c_m(\zeta)+i s_m(\zeta) = \int_0^\infty K^m e^{-K^3+iK\zeta}dK.
\end{eqnarray}
In Eqs.~(\ref{eq:mooresaff}), (\ref{eq:wy}) and (\ref{eq:msf}),
the integer $m+1$ corresponds to the multipolar order of the line
source of waves as discussed in~\cite{Machicoane2015}: $m=0$
corresponds to a monopolar source and $m=1$ to a dipolar source.
Our wavemaker produces a large-scale wave with a zero
instantaneous net mass flux (because of an integer number of wavelengths) which
suggests to consider the dipolar
case $m=1$. However, it is worth noting that Jouve and
Ogilvie~\cite{Jouve2014} consider the case $m=0$ in
their work which leads to a successful description of the spatial dependence of
the attractor amplitude in their numerical simulations. The multipolar order of
the virtual source to be considered here is therefore an open
question.

Noting $L_a$ the length of the unwrapped inviscid attractor and
$L_0$ the distance between the virtual source and the focusing
point, the balance between the viscous spreading of the wave
between $\xi=L_0$ and $\xi=L_0+L_a$ with lengthscales increasing
as $\xi^{1/3}\ell^{2/3}$ and the focusing reflection leads to the
compatibility relation $L_0=L_a/(\gamma^3-1)$. In the experiments,
the attractor length $L_a$ is $214.2$~cm such that $L_0\simeq
20.5$~cm (see Fig.~\ref{fig:cavite_attract}). This eventually
leads to the following relation for the transverse length scale of the
attractor as a function of along-attractor coordinate
$s$~\cite{Grisouard2008,Jouve2014,Brouzet2017}
\begin{eqnarray}\label{eq:spreading}
\frac{\delta}{L_a}=\left(\frac{\ell}{L_a}\right)^{2/3}\left(\frac{s}{L_a}
+ \frac{1}{\gamma^3-1}\right)^{1/3}.
\end{eqnarray}
Here, $s=\xi-L_0$ is the distance along the unwrapped attractor
starting from the focusing reflection.

The relevance of the scaling law~(\ref{eq:spreading}) has been
first tested numerically for gravity waves in a stratified fluid
by Grisouard and coworkers~\cite{Grisouard2008} who indeed
observed power law behaviors but with exponents departing from
$1/3$ by $-25\%$ and $+40\%$ in the two reported configurations.
In numerical simulations of a rotating tilted square, Jouve and
Ogilvie~\cite{Jouve2014} confirmed more clearly the predicted
scaling laws for the attractor width $(\xi/\ell)^{1/3}$ and the
velocity amplitude $(\xi/\ell)^{-1/3}$ (case $m=0$) with the
position along the attractor as well as the relevance of the
Moore-and-Saffman transverse structure of the beam. From an
experimental point of view, Brouzet \textit{et
al.}~\cite{Brouzet2017} studied the evolution with time of the
wavelength in the attractor during its transient growth and decay
phases in a linearly stratified fluid. These data revealed a
decrease toward a steady state value during the forced growth and
a further decrease during the free decay. Comparable results were
previously reported by Hazewinkel and
coworkers~\cite{Hazewinkel2008}.

Jouve and Ogilvie~\cite{Jouve2014} also studied the non-linear
regime of the attractor of inertial waves and showed the emergence
of a local instability close to each focusing point. This
instability transfers the energy of the wave attractor toward two
subharmonic waves with their frequencies in triadic resonance with
the primary wave frequency. This scenario is consistent with the
one reported in internal gravity wave attractor experiments by
Scolan \textit{et al.}~\cite{Scolan2013}. These local subharmonic
instabilities of an internal wave attractor are actually very
similar to instabilities of plane wave beams observed
experimentally in rotating~\cite{Bordes2012} and
stratified~\cite{Bourget2013} fluids. In this context, it has been
shown~\cite{Bourget2014,Karimi2014} that one should take into account the
finite width of the wave beam,
i.e. the small number of wavelengths contained in the transverse extension of
the wave attractor (typically 1), in order to correctly predict
the growth rate of the secondary waves generated by the
instability.

In this context, one can highlight a remarkable feature of the
instability of the experimental plane inertial wave reported by
Bordes \textit{et al.}~\cite{Bordes2012} compared to all other
mentioned instabilities: in the temporal spectrum, two wide bumps
are observed, centered around two frequencies in triadic resonance
with the primary wave frequency. This is in strong contrast with
the internal gravity waves experiments with either a plane
wave~\cite{Bourget2013} or an attractor~\cite{Scolan2013} but also
with the inertial wave attractor simulations~\cite{Jouve2014} for
which the instability is very selective in terms of secondary wave
frequencies. The physical origin of this specificity of
experimental inertial waves remains unclear up to now.

Finally, in~\cite{Jouve2014} and~\cite{Brouzet2017} in which
subharmonic triadic instabilities of numerical inertial and
experimental internal wave attractors are reported, a thickening
of the attractor beam is reported, as the forcing amplitude is
increased. This thickening can be understood qualitatively as the
consequence of the extraction of energy from the attractor at
frequency $\sigma_0$ by the triadic instability which acts as an
effective turbulent dissipation: the instability then naturally
produces an attractor with a smaller relative amplitude and larger
transverse scales as the forcing amplitude is increased.

\section{Experimental results}

\subsection{Linear regime}

In our experiments, energy is injected at frequency $\sigma_0 =
0.85\times 2\Omega$ by the wave generator. In order to uncover the
frequency content of the flow, we compute for each experiment the
temporal power spectral density of the velocity field as
\begin{eqnarray}\label{eq:psd}
E(\sigma)=\frac{4\pi}{T}\langle|\tilde{u}_j(x,z,\sigma)|^2\rangle,
\end{eqnarray}
where
\begin{eqnarray}
\tilde{u}_j(x,z,\sigma)=\frac{1}{2\pi}\int_0^T
u_j(x,y_0,z,t)e^{-i\sigma t} \,dt
\end{eqnarray}
is the temporal Fourier transform of $u_j(x,y_0,z,t)$ with
$j=x,z$, the angular brackets denote the spatial average over the
measurement plane, $T$ is the acquisition duration and $y_0 =
L_y/3$.

\begin{figure}
    \centerline{\includegraphics[width=9.5cm]{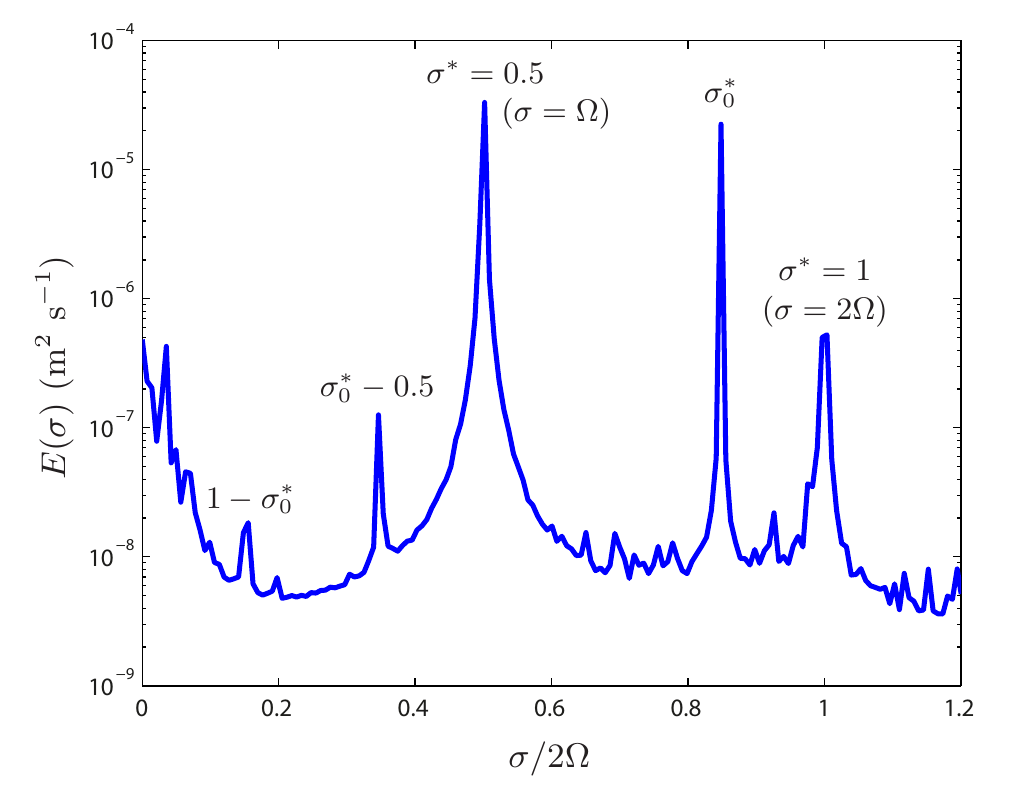}}
    \caption{Temporal power spectral density $E(\sigma)$
    (Eq.~\ref{eq:psd}) as a function of the normalized frequency
    $\sigma^*=\sigma/2\Omega$ for $A=0.09$~mm, $\lambda_f=52.4$~cm and
    $\Omega=18$~rpm. We have highlighted five peaks at frequencies
    $\sigma^*_0=\sigma_0/2\Omega$, $\sigma^*=0.5$ (i.e.
    $\sigma=\Omega$), $\sigma^*=1$ (i.e. $\sigma=2\Omega$),
    $\sigma^*=\sigma_0^*-0.5$ and $\sigma^*=1-\sigma_0^*$.}\label{fig:spectrefreq0_09}
\end{figure}

\begin{figure}
    \centerline{\includegraphics[width=14cm]{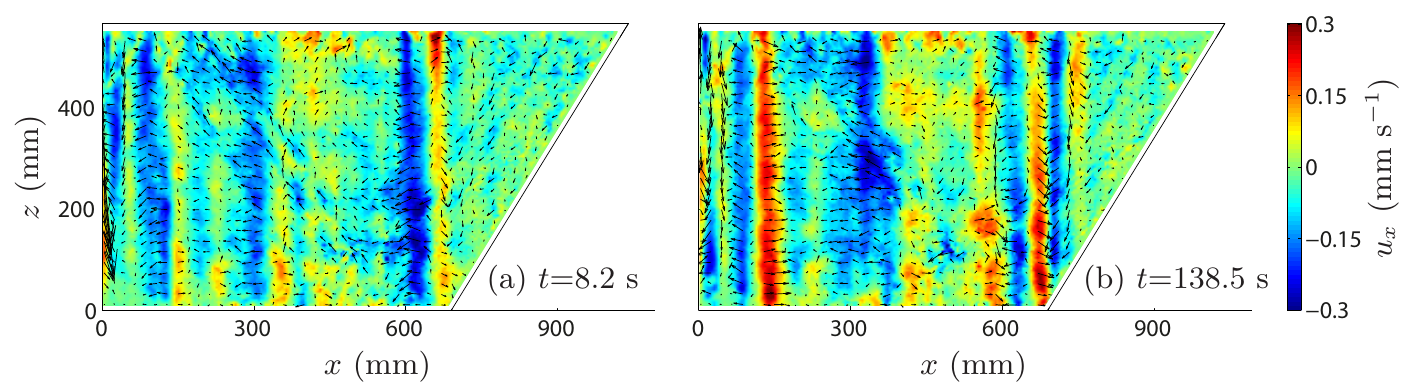}}
    \caption{Snapshots of the velocity field after a temporal
    moving average over a time window of 8 wavemaker periods (i.e. a cutoff frequency of
    $\sigma^* \simeq 0.11$) in the vertical plane $y=y_0= L_y/3$ for $\lambda_f=52.4$~cm and the lowest forcing
    amplitude $A=0.09$~mm at $\Omega=18$~rpm.}\label{fig:BF}
\end{figure}

In Fig.~\ref{fig:spectrefreq0_09}, we report the temporal energy
spectrum $E(\sigma)$ for the forcing wavelength
$\lambda_f=52.4$~cm and the lowest forcing amplitude $A=0.09$~mm
at $\Omega=18$~rpm as a function of the normalized frequency
$\sigma^*=\sigma/2\Omega$. This spectrum is mainly composed of a
sharp peak at the forcing frequency $\sigma_0$ as well as a
secondary but energetic peak at the frequency of the rotating
platform $\sigma=\Omega$. The latter corresponds to a flow excited
by the Earth rotation which induces a precession of the rotating
platform~(see Refs.~\cite{Boisson2012,Triana2012} and references
therein). In the case of a spherical or ellipsoidal cavity under
precession, it is known as the tilt-over flow and has been
extensively studied due to its relevance in
astrophysics~\cite{Kida2011}. One can also note the presence of
energy at low frequencies $\sigma/2\Omega<0.15$. In a
previous work of two authors of this paper using the same rotating
platform~\cite{Bordes2012}, it was shown that the energy peak at
zero frequency (with a tail extending up to $\sigma^*\simeq 0.15$)
was already present when the flow forcing is off, suggesting that
the low frequency spectral component is possibly the consequence
of thermal convection in the water tank. The observation of the
velocity field temporally smoothed over a large time-scale
actually revealed the presence of columns which are slowly
drifting in the water tank, resembling the columns observed in
rotating thermal convection experiments~\cite{Sakai1997,King2012}.
Here, we observe the same kind of vertical columns, dominated by
horizontal velocities, in the low-pass frequency filtered velocity
field (two snapshots are shown in Fig.~\ref{fig:BF}) which could
possibly be the consequence of thermal convection. Nevertheless,
even if no clear signs are observed here, one should let open the
possibility that part of the energy present at very low
frequencies in our experiments could be related to non-linearities
---such as steady streaming~\cite{Sauret2010,Bordes2012b} or
Stokes drift~\cite{Sutherland2006}--- affecting the flow motions
at the forcing frequency $\sigma_0$ or at the platform frequency
$\Omega$. Other weakly energetic peaks are also present in the
spectrum corresponding to the first harmonic of the tilt-over flow
($\sigma^*=1$), to interactions between the forcing and the
tilt-over flow at $\sigma^*=\sigma_0^*-0.5\simeq 0.35$ and to
interactions between the forcing and the first harmonic of the
tilt-over flow at $\sigma^*=1-\sigma_0^*\simeq 0.15$.

\begin{figure}
    \centerline{\includegraphics[width=0.85\textwidth]{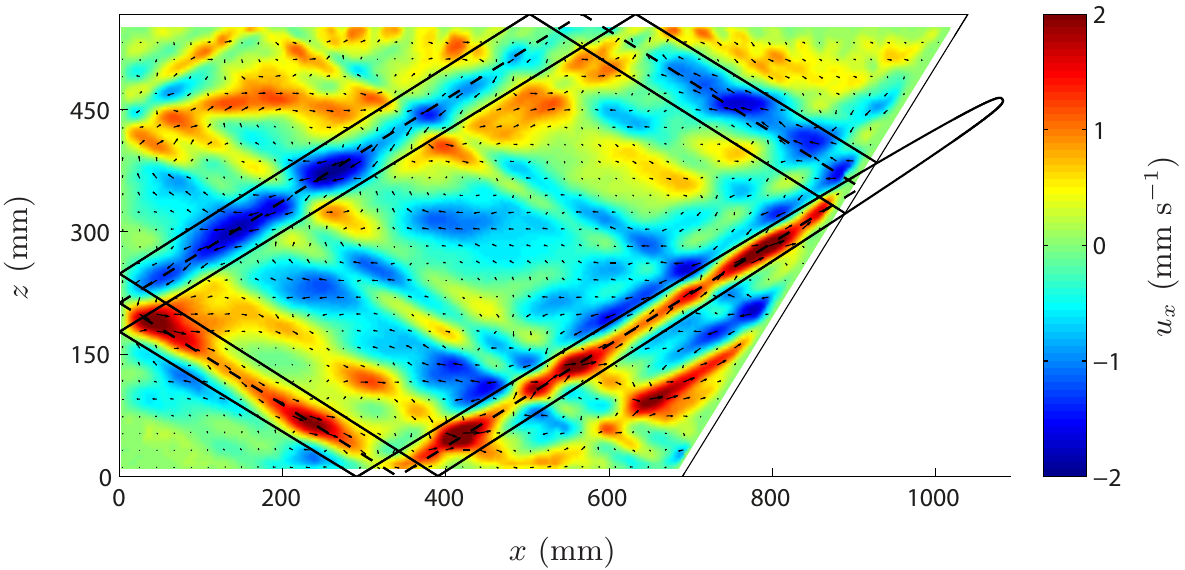}}
    \caption{Attractor in the linear regime: Snapshot of the velocity field in
    the
    vertical plane $y=y_0= L_y/3$ Fourier filtered at the forcing frequency
    $\sigma_0$ for $\lambda_f=52.4$~cm and the lowest forcing
    amplitude $A=0.09$~mm at $\Omega=18$~rpm. A sketch of the
    theoretical attractor is superimposed to the experimental field:
    the dashed line shows the inviscid skeleton and the two solid
    lines delineate the width at mid-height of the viscous beam
    longitudinal velocity amplitude~(Eq.~\ref{eq:mooresaff}).}\label{fig:vmot}
\end{figure}

Overall, the flow produced by the wave generator with the lowest
forcing amplitude at $\Omega=18$~rpm seems to be in the linear
regime. In order to discard the tilt-over and the low frequency
flow components, we Fourier filter the velocity field at the
forcing frequency $\sigma_0$. A snapshot of the corresponding
field is reported in Fig.~\ref{fig:vmot} to which is superimposed
the width at mid-amplitude of the theoretical
attractor~(\ref{eq:mooresaff}). The velocity field reveals a
concentration of energy along the theoretical attractor in good
agreement with the theory. This concentration is however only
partial since one can see other wave beams tilted by the angle
$\theta$ outside of the region where the theoretical attractor is
expected: in Fig.~\ref{fig:vmot}, the velocity magnitude in
the attractor beam (from $1$~mm/s to $2$~mm/s) is actually only 3
to 7 times larger than the forcing velocity magnitude $U_f \simeq
0.27$~mm/s. As a consequence, since the wavemaker injects energy
over the whole width of the water tank, we naturally find wave
beams with a non-negligible amplitude outside of the attractor
region.

\begin{figure}
    \centerline{\includegraphics[width=13cm]{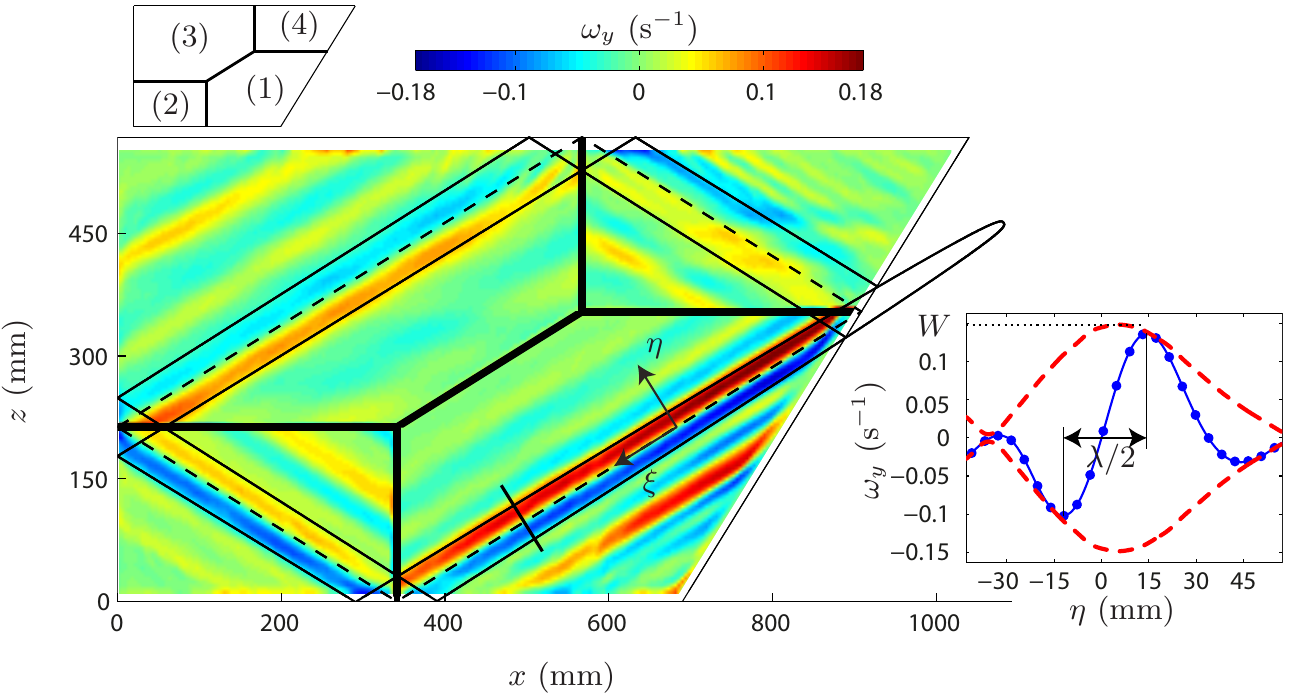}}
    \caption{Attractor in the linear regime: Snapshot of the out-of-plane
    vorticity
    of the Hilbert filtered field at the forcing frequency $\sigma_0$
    for $A=0.09$~mm, $\lambda_f=52.4$~cm and $\Omega=18$~rpm. The
    reported field is actually a combination of four regions in which
    different wavevector quadrant have been selected by the Hilbert
    filtering in agreement with the direction expected for the
    wavevector in each attractor branch (cf.
    Fig.~\ref{fig:cavite_attract}): In region (1) we keep the
    wavevector quadrant ($k_x>0, k_z<0$), in (2) ($k_x>0, k_z>0$), in
    (3) ($k_x<0, k_z>0$) and in (4) ($k_x<0,k_z<0$). A sketch of the
    theoretical attractor is superimposed (same layout as in
    Fig.~\ref{fig:vmot}). In inset: experimental transverse profile
    (blue line with data markers) of the $y$-component of the
    vorticity $\omega_y(\xi=s+L_0,\eta,\varphi)$ of the $\sigma_0$ and
    $(k_x>0,k_z<0)$ Hilbert-filtered field (region 1). It is taken at
    coordinate $s=49.7$~cm along the attractor axis (corresponding to
    the straight line in the snapshot) and at a given arbitrary phase.
    We also report in the inset the experimental wave beam envelope
    (red dashed line) computed from the experimental transverse
    profiles as $\omega_{y,0}=\sqrt{2\langle
    \omega_y(\xi,\eta,\varphi)^2\rangle_\varphi}$ where $\langle\,
    \rangle_\varphi$ stands for the average on the phase $\varphi \in
    [0,~2\pi]$.}\label{fig:branche1}
\end{figure}

To further compare the experimental flow at frequency $\sigma_0$
and the theoretical attractor, we study its transverse profile as
a function of the longitudinal coordinate $s=\xi-L_0$ along the
inviscid attractor. To do so, we notice that each of the four
branches of the attractor has its wavevector in a different
quadrant of the wave vector plane $(k_x,k_z)$ (the wave vector is
aligned with the phase velocity $\bf c_\varphi$, see
Fig.~\ref{fig:cavite_attract}). We perform a Hilbert filtering of
the velocity field (see Ref.~\cite{Mercier2008} for details). It
consists in computing the temporal Fourier transform of the raw
velocity field, band-pass filtering the result around the
frequency of interest $\sigma_0$ (keeping only positive
frequencies), and computing the inverse Fourier transform. We then
take the two-dimensional (2D) spatial Fourier transform of the
resulting complex field relative to $x$ and $z$, put to zero the
values of the resulting field except in the wavevector quadrant of
interest, and finally compute the inverse 2D Fourier transform in
space. Taking twice the real part of the result eventually
provides the velocity field of the waves at frequency~$\sigma_0$
and with their wavevector in a given quadrant~\cite{Mercier2008}.
We finally perform a temporal phase average
at~$\sigma_0$~\cite{phaseaverage}. As an illustration, we report
in Fig.~\ref{fig:branche1} a snapshot of the field resulting from
the Hilbert filtering: this snapshot is divided in four regions,
each corresponding to a given theoretical attractor branch and to
a Hilbert filtering selecting the wavevector quadrant of the
theoretical attractor branch. One can note the presence of a few
wave beams outside of the theoretical attractor which reveals that
the focusing of the energy injected by the forcing in the
attractor although clear is only partial.

In the inset of Fig.~\ref{fig:branche1}, we report a transverse
profile (along $\eta$) of the out-of-plane vorticity component
$\omega_y(\xi=s+L_0,\eta,\varphi)$ corresponding to the $\sigma_0$
and $(k_x>0,k_z<0)$ Hilbert filtered velocity field. This
transverse profile is taken at coordinate $s=49.7$~cm along the
attractor axis (corresponding to the solid line in
Fig.~\ref{fig:branche1}) at a given arbitrary phase, still for
$\lambda_f=52.4$~cm and the lowest forcing amplitude $A=0.09$~mm
at $\Omega=18$~rpm. We also report the corresponding experimental
wave beam envelope $\omega_{y,0}(\xi,\eta)=\sqrt{2\langle
\omega_y(\xi,\eta,\varphi)^2\rangle_\varphi}$ where $\langle\,
\rangle_\varphi$ stands for the average on the phase
$\varphi\in[0,~2\pi]$. From such curves, we measure as a function
of the longitudinal position $\xi$, the beam vorticity amplitude
$W(\xi)=\textrm{max}_\eta[\omega_{y,0}(\xi,\eta)]$ as well as the
wavelength $\lambda(\xi)$ estimated as the mean value over
$\varphi$ of twice the transverse distance between the maximum and
minimum of the vorticity profile $\omega_y(\xi,\eta,\varphi)$.

\begin{figure}
    \centerline{\includegraphics[width=10cm]{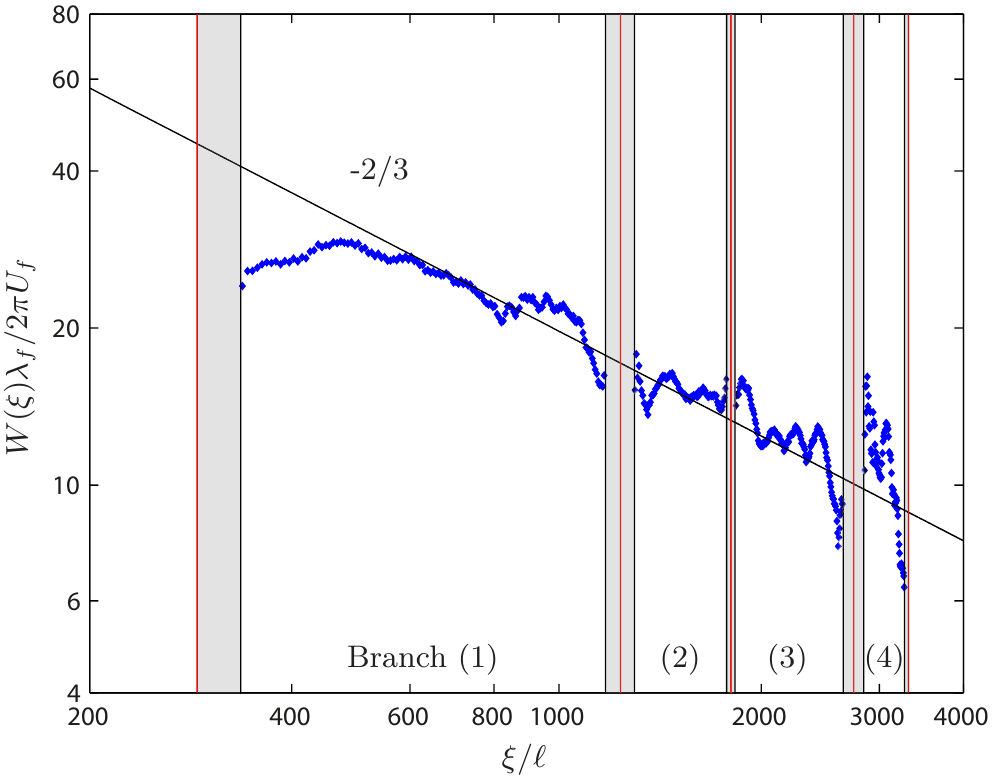}}
\caption{Vorticity amplitude $W(\xi)$ of the attractor as a
function of the coordinate $\xi=s+L_0$ for $A=0.09$~mm,
$\lambda_f=52.4$~cm and $\Omega=18$~rpm. $W(\xi)$ is normalized by
the ``forcing vorticity'' $2\pi U_f/\lambda_f$ and the position
$\xi$ by the viscous lengthscale $\ell$ given by Eq.~(\ref{eq:ell}). Vertical
red lines show the reflections on the cavity walls and grey
regions delineate zones in which PIV measurement is not possible.
Following Eq.~(\ref{eq:wy}), a power law with an exponent $-2/3$
is also shown as a guide for the eyes.}\label{fig:powerlaw}
\end{figure}

In Fig.~\ref{fig:powerlaw}, we report the vorticity amplitude
$W(\xi)$ for $A=0.09$~mm, $\lambda_f=52.4$~cm and $\Omega=18$~rpm
as a function of coordinate $\xi=s+L_0$ along the unwrapped
theoretical beam emitted by the virtual source. Data are missing
on five portions of $\xi$ corresponding to the regions where the
velocity field cannot be measured by PIV close to the reflections
on the cavity walls. The attractor amplitude $W(\xi)$ shows
significant oscillations that are due to interferences of the wave
attractor with the additional inertial waves at $\sigma_0$ present
in the cavity (see Fig.~\ref{fig:vmot}) as well as to
interferences between two branches of the attractor close to a
reflection. Nevertheless, one can observe a good agreement between
the data and a power law of exponent $-2/3$ in agreement with the
scaling predicted by the theory~(\ref{eq:wy}) for a monopolar
source of waves $m=0$. The observation of this $-2/3$ spatial
decay exponent is consistent with the numerical data reported by
Jouve and Ogilvie~\cite{Jouve2014}. It shows that the multipolar
order of the virtual point source to be considered in the
attractor model is $m=0$ (monopolar source) and seems largely
independent of the way energy is injected into the system.

\begin{figure}
    \centerline{\includegraphics[width=10cm]{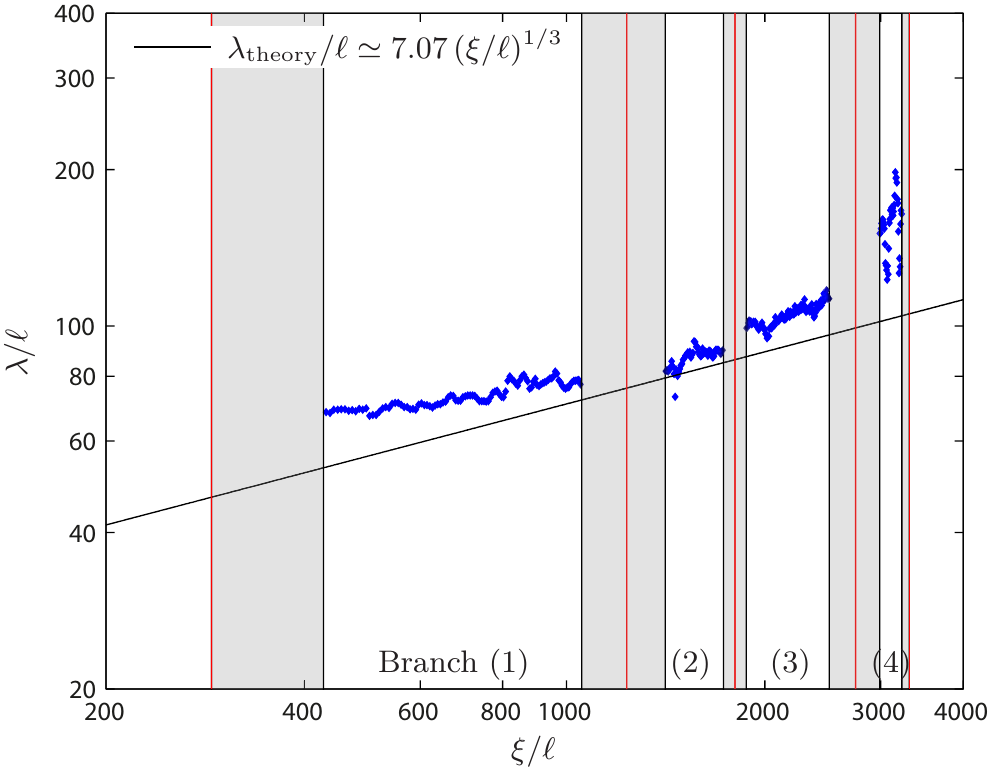}}
\caption{Normalized wavelength $\lambda/\ell$ in the attractor
beam as a function of the normalized distance from the virtual
source $\xi/\ell$ for $A=0.09$~mm, $\lambda_f=52.4$~cm and
$\Omega=18$~rpm. Vertical red lines show the reflections on the
cavity walls and grey regions delineate zones in which boundary
effect prevents measurements. The solid line shows the theoretical
predictions, in $(\xi/\ell)^{1/3}$ and with no adjustable
parameter, for the wavelength $\lambda$.\label{fig:lambda}}
\end{figure}

In Fig.~\ref{fig:lambda}, we show the corresponding evolution with
$\xi$ of the wavelength $\lambda$ in the attractor, normalized by
the viscous lengthscale $\ell$~(Eq.~\ref{eq:ell}). As in
Fig.~\ref{fig:powerlaw}, data are missing around the reflections
on the cavity walls. The excluded ranges of $\xi/\ell$ are larger
because estimates of the attractor transverse lengthscales are
prevented when approaching a wall at distance of the order of
these lengthscales ($\sim 100-200 \times \ell$). We also report in
Fig.~\ref{fig:lambda} the theoretical prediction for
$\lambda/\ell$ according to Eqs.~(\ref{eq:wy}-\ref{eq:msf}). We
emphasize that this prediction is a power law in
$(\xi/\ell)^{1/3}$ with a prefactor theoretically prescribed by
the Moore-and-Saffman functions. One sees that, despite the fact
that the power law behavior is not clearly observed in the data,
the theory provides correct estimates for $\lambda$. The
wavelength is found here always slightly larger than the
theoretical prediction. Such a tendency is identical to the one
reported for experimental gravity waves
attractor~\cite{Brouzet2017}. In this work as well as
in~\cite{Beckebanze2018}, it is proposed that the additional
dissipation due to the viscous friction on the vertical walls of
the cavity ($y=0$ and $y=L_y$) leads to an attractor larger than
in the 2D theory (invariant in the $y$~direction) by modifying the
balance between energy focusing and viscous dissipation. One
can finally highlight that in both Figs.~\ref{fig:powerlaw}
and~\ref{fig:lambda}, the experimental data in the fourth branch
of the attractor are particularly noisy and also significantly
departing from the theoretical scaling law. This could be
understood by the fact that the fourth branch is the weaker in
magnitude (see Eq.~\ref{eq:mooresaff}) whereas at the same time it
is located where the original wave produced by the wavemaker is
the strongest. The experimental data in the fourth branch of the
attractor are therefore probably strongly affected by
interferences between the wave in the attractor and the original
wave produced by the wavemaker.

\subsection{Non-linear regime}

\begin{figure}
    \centerline{\includegraphics[width=10cm]{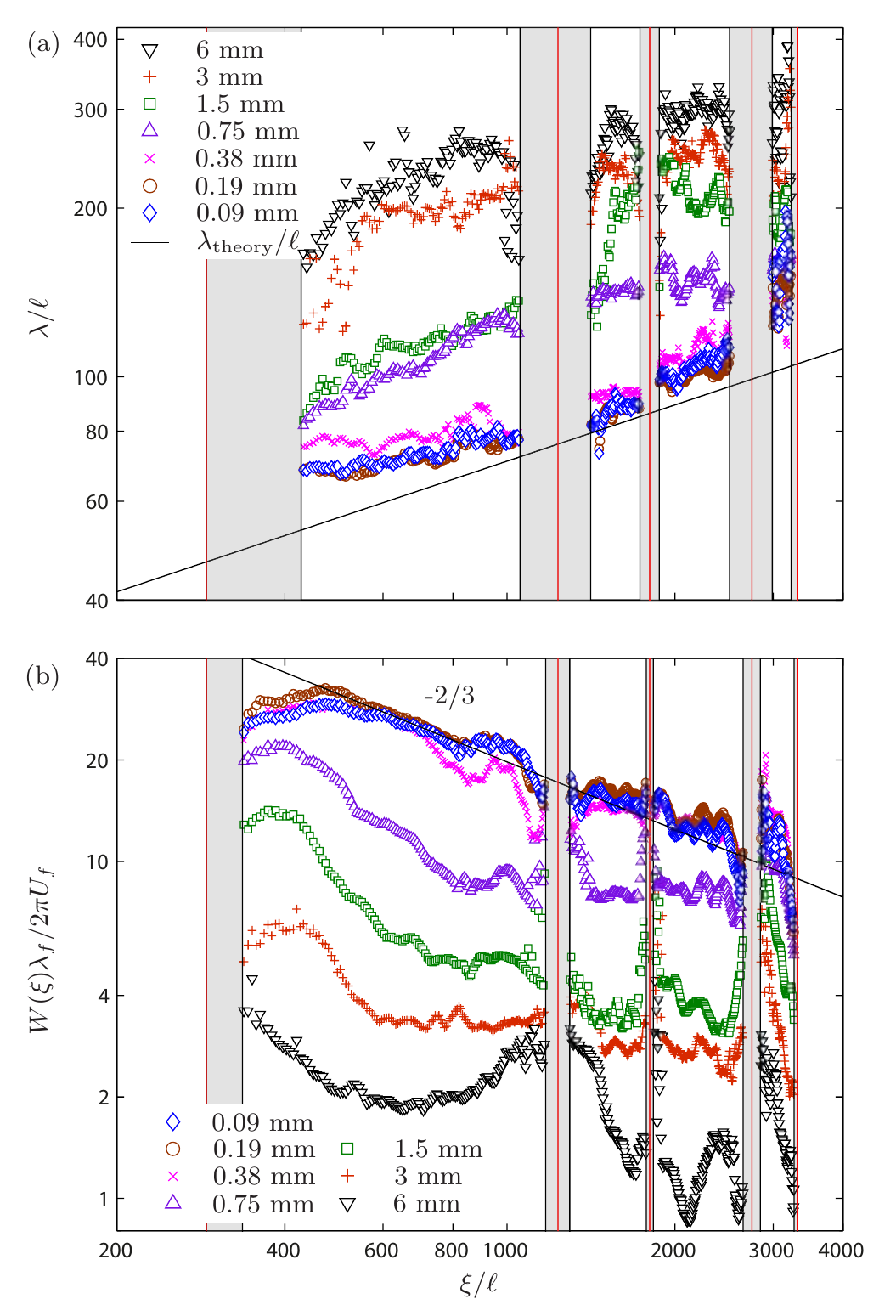}}
    \caption{(a) Normalized wavelength $\lambda/\ell$ and (b) vorticity
    amplitude $W(\xi) \lambda_f/ 2\pi U_f$ as a function of
        the normalized distance from the virtual source $\xi/\ell$ for each
        forcing amplitude $A$ at $\Omega=18$~rpm and $\lambda_f=52.4$~cm. For
        both figures, vertical
        red lines show the reflections of the theoretical attractor on the
        cavity walls and grey regions zones in which boundary
        effect prevents measurements. In (a), the solid line shows the
        theoretical power law, in $(\xi/\ell)^{1/3}$ and with no
        adjustable parameter, for the linear attractor. In (b), a power law
        with an exponent $-2/3$ is shown, corresponding to the theoretical
        linear attractor.\label{fig:alllaw}}
\end{figure}

\begin{figure}
    \centerline{\includegraphics[width=11cm]{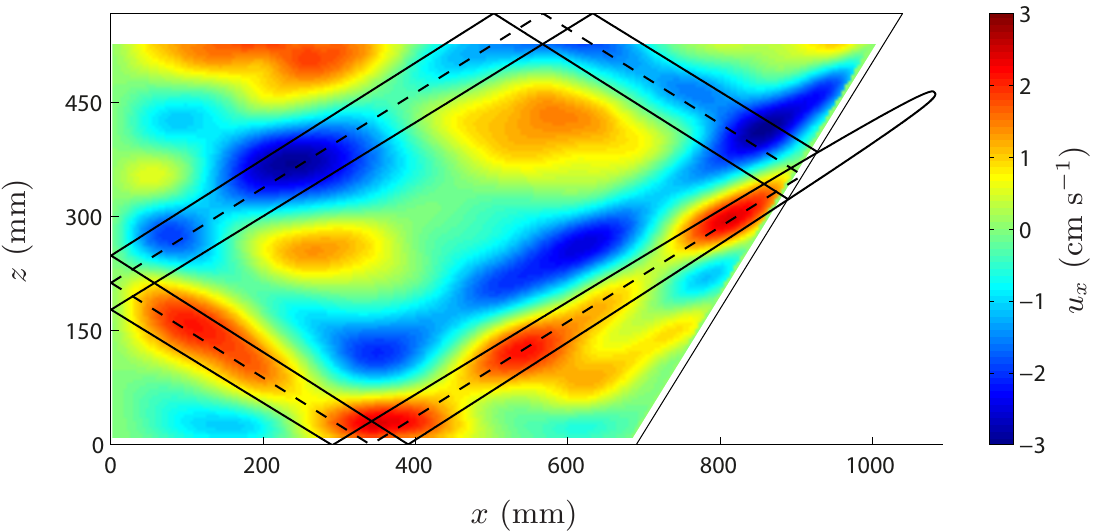}}
    \caption{Attractor in the nonlinear regime: Snapshot of the velocity field
    in
    the vertical plane $y=y_0= L_y/3$ Fourier filtered at the forcing
    frequency $\sigma_0$ for $A=3.00$~mm, $\lambda_f=52.4$~cm and
    $\Omega=18$~rpm. As in Fig.~\ref{fig:vmot}, a sketch of the
    theoretical linear attractor is superimposed to the experimental
    field.}\label{fig:vmotGA}
\end{figure}

We now repeat the previous analysis for increasing forcing
amplitude $A$. In Fig.~\ref{fig:alllaw}, we report (a) the
wavelength and (b) the out-of-plane vorticity amplitude normalized
by the ``forcing vorticity'' $2\pi U_f/\lambda_f$ as a function of
the coordinate $\xi$, for all forcing amplitudes at
$\Omega=18$~rpm and $\lambda_f=52.4$~cm. One can see that the
wavelength and the normalized vorticity amplitude are nearly
identical for the three lowest forcing amplitudes indicating that
the flow is in the linear regime. For larger values of the forcing
amplitude $A$, the transverse (cross-beam) wavelength of the beam
increases whereas the normalized beam vorticity decreases with
$A$, indicating the emergence of non-linear effects.
Figure~\ref{fig:vmotGA}, showing a snapshot of the velocity field
Fourier filtered at $\sigma_0$ for $A=3.00$~mm at $\Omega=18$~rpm
and $\lambda_f=52.4$~cm, provides a direct illustration of the
attractor thickening when increasing $A$. In this figure, one
still observes a concentration of energy around the theoretical
attractor but this concentration is clearly less pronounced than
for low $A$ (see Fig.~\ref{fig:vmot}).

\begin{figure}
    \centerline{\includegraphics[width=8cm]{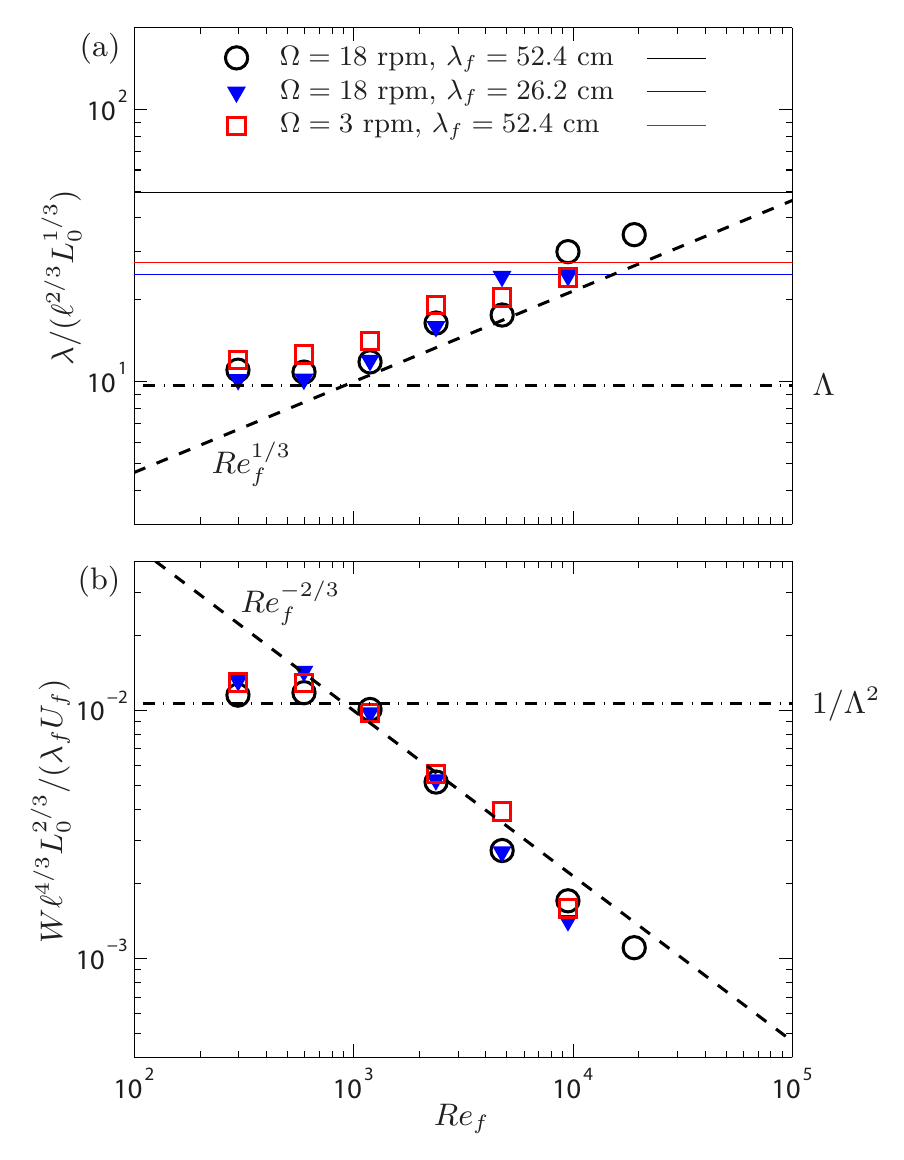}}
    \caption{(a) Normalized wavelength $\lambda/(\ell^{2/3}L_0^{1/3})$
    and (b) normalized vorticity amplitude
    $W\ell^{4/3}L_0^{2/3}/(\lambda_f U_f)$ averaged over the first
    branch of the attractor (the one following the focusing
    reflection) as a function of the forcing Reynolds number
    $Re_f=U_f\lambda_f/\nu$. Square symbols correspond to experiments
    at ($\Omega=3$~rpm, $\lambda_f=52.4$~cm), triangles to
    ($\Omega=18$~rpm, $\lambda_f=26.2$~cm) and circles to
    ($\Omega=18$~rpm, $\lambda_f=52.4$~cm). The dashed lines in (a) and
    in (b) show respectively the scaling laws
    $\lambda/(\ell^{2/3}L_0^{1/3})=Re_f^{1/3}$ and
    $W\ell^{4/3}L_0^{2/3}/(\lambda_f U_f)=Re_f^{-2/3}$ predicted when
    replacing the fluid viscosity by the turbulent viscosity
    $\nu_t=U_f\lambda_f$ in the viscous length $\ell$ in the linear
    attractor model. In (a), the horizontal dashed-dotted line shows the
    theoretical value $\Lambda$ for $\lambda/(\ell^{2/3}L_0^{1/3})$ predicted by
    the linear attractor model described in section~\ref{sec:theo} (average of the theoretical value over the first branch).
    In (b), the horizontal dashed-dotted line shows the corresponding numerical
    value $1/\Lambda^2$ which stands as an estimate for 
    $W\ell^{4/3}L_0^{2/3}/(\lambda_f U_f)$ (see main text).
    The solid horizontal lines, black, red and blue, show the theoretical
    wavelength $\lambda_f/\gamma$ of the beam excited by the wavemaker after one
    reflection on the sloping wall for the three experimental
    configurations, ($\Omega=18$~rpm, $\lambda_f=52.4$~cm), ($\Omega=3$~rpm,
    $\lambda_f=52.4$~cm) and ($\Omega=18$~rpm,
    $\lambda_f=26.2$~cm) respectively.}\label{fig:lambdaA}
\end{figure}

\begin{figure}
    \centerline{\includegraphics[width=13cm]{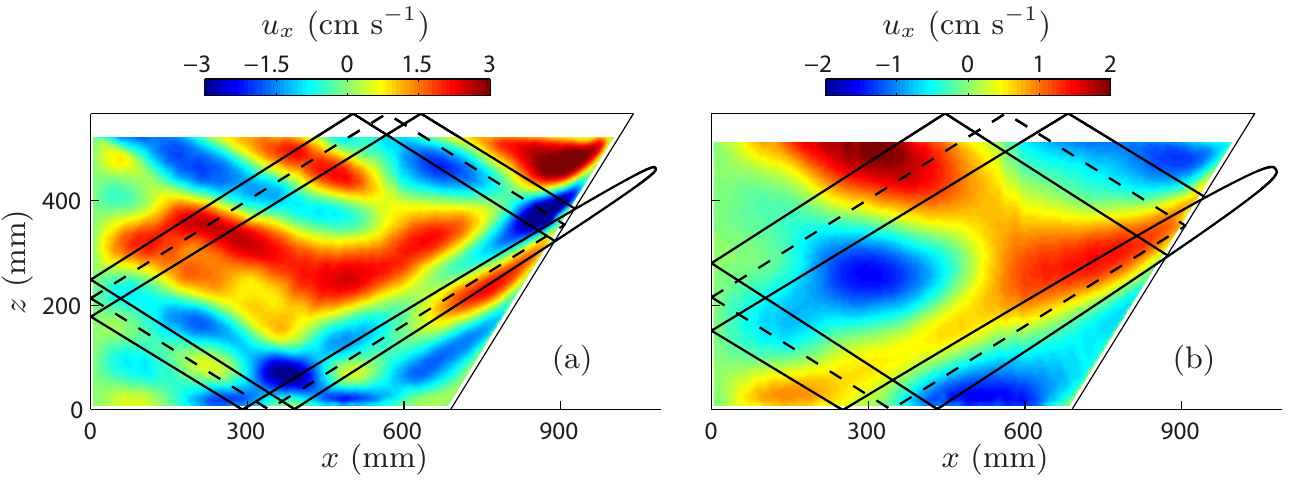}}
\caption{Experiments without attractor: Snapshots of two velocity
fields in the vertical plane $y=y_0= L_y/3$ at $Re_f=9\,500$,
Fourier filtered at $\sigma_0$ for (a) $A=6$~mm,
$\lambda_f=26.2$~cm and $\Omega=18$~rpm and (b) $A=18$~mm,
$\lambda_f=52.4$~cm and $\Omega=3$~rpm. No attractor is observed
in these fields because the wavelength of the theoretical
non-linear attractor is larger than the forcing wavelength. The
absence of attractor is therefore a combined effect of the
non-linearities and the forcing. As in Fig.~\ref{fig:vmot}, a
sketch of the corresponding theoretical linear attractor is
superimposed to each experimental fields.}\label{fig:snap3rpm}
\end{figure}

In Fig.~\ref{fig:lambdaA}, we report, as a function of the forcing
Reynolds number $Re_f=U_f\lambda_f/\nu$, (a) the wavelength
$\lambda$ and (b) the vorticity amplitude $W$ averaged over the
first branch of the attractor (the one following the focusing
reflection). Three data series are reported here for
($\Omega=3$~rpm, $\lambda_f=52.4$~cm), ($\Omega=18$~rpm,
$\lambda_f=52.4$~cm) and ($\Omega=18$~rpm, $\lambda_f=26.2$~cm).
In Fig.~\ref{fig:lambdaA}(a), the wavelength is normalized by
$\ell^{2/3}L_0^{1/3}$ accounting for the dependence predicted by
the linear attractor theory~(\ref{eq:delta}). We first note that
the three data series collapse on a master curve. This
suggests that the forcing Reynolds number $Re_f$ is, to the first
order, the parameter controlling the non-linear evolution of the
attractor. The normalized attractor wavelength
$\lambda/\ell^{2/3}L_0^{1/3}$ is close to the value predicted by
the linear theory $\Lambda \simeq 9.68$ for $Re_f \lesssim 1000$
and increases at larger $Re_f$ as already observed in
Fig.~\ref{fig:alllaw}(a) ($\Lambda$ is the average of the
theoretical prediction over the attractor first branch; it is
shown by the horizontal dashed-dotted line). One can however note
that the normalized wavelength saturates at the larger Reynolds
numbers for the series at ($\Omega=3$~rpm, $\lambda_f=52.4$~cm)
and ($\Omega=18$~rpm, $\lambda_f=26.2$~cm). This saturation is
easy to understand: the wavelength $\lambda$ cannot be larger than
the one of the forcing, since energy can only be transferred to
smaller scales, via the focusing reflections. In
Fig.~\ref{fig:lambdaA}(a), the horizontal solid lines (black, red
and blue) show the normalized wavelength $\lambda_f/\gamma$ for
the three series of experiments. This wavelength theoretically
correspond to the wave excited by the wavemaker after one
reflection on the sloping wall. It stands as an upper limit for
the wavelength found in the first branch of the attractor. For the
largest forcing amplitudes at ($\Omega=3$~rpm,
$\lambda_f=52.4$~cm) and ($\Omega=18$~rpm, $\lambda_f=26.2$~cm),
$\lambda$ approaches this limit suggesting that almost no energy
concentration in the attractor is observed. This is confirmed by
the direct observation in Fig.~\ref{fig:snap3rpm} of two
corresponding velocity fields at $Re_f\simeq 9\,500$ in which one
typically sees the wave excited by the wavemaker reflecting on the
sloping wall: the forcing wavelength being smaller than the
theoretical wavelength expected for the non-linear attractor,
energy cannot be supplied to the latter by the forcing. In
comparison, for the data series at ($\Omega=18$~rpm,
$\lambda_f=52.4$~cm), the wavelength for the largest Reynolds
number $Re_f\simeq 19\,000$ is still significantly lower than the
excited wave original wavelength after one reflection
$\lambda_f/\gamma$, revealing a greater robustness of the energy
concentration in an attractor to the increase of Reynolds number
when the rotation or the injection scale are larger, i.e. when the
forcing Rossby number $Ro_f=U_f/2\Omega \lambda_f$ is lower.

As mentioned in Sec.~\ref{sec:theo}, we expect that in an inviscid
fluid the product of the velocity times the wavelength of an
inertial wave is conserved during the reflection on a tilted wall.
A tentative scaling law for the vorticity amplitude of the linear
attractor is therefore $W_t=U_f \lambda_f/(\ell^{2/3}L_0^{1/3})^2$
where $U_f$ and $\lambda_f$ are characteristic of the wave
initially forced by the wavemaker and $\ell^{2/3}L_0^{1/3}$ is the
theoretical scaling for the linear attractor wavelength. In
Fig.~\ref{fig:lambdaA}(b), we report the vorticity amplitude $W$
normalized by $W_t$ as a function of $Re_f$. This normalization
collapses the three data series on a master curve which
illustrates that $W_t$ catches the physics of the attractor
amplitude in the linear and non-linear regime. We verify that this
normalized vorticity $W/W_t$ is first constant at low forcing
Reynolds number $Re_f \lesssim 1\,000$ confirming the linear
regime of the flow. A tentative estimate for the normalized
attractor vorticity $W/W_t$ in the linear regime could be made by
considering the theoretical value predicted by the linear model
for the attractor wavelength, i.e. $W/W_t=1/\Lambda^2\simeq
1.07\times 10^{-2}$ ($\Lambda\simeq 9.68$ is the average over the first branch 
of the theoretical normalized attractor wavelength). This
prediction for $W/W_t$ is reported with a horizontal dashed-dotted
line in Fig.~\ref{fig:lambdaA}(b): one sees that it indeed
provides a reasonable estimate of the attractor vorticity in the
linear regime. This behavior is consistent with the fact that the
wavelength matches the linear theory in Fig.~\ref{fig:lambdaA}(a)
for the same Reynolds number range. At larger $Re_f$, the ratio
$W/W_t$ decreases with $Re_f$ revealing again the emergence of
non-linearities.

\begin{figure}
    \centerline{\includegraphics[width=8cm]{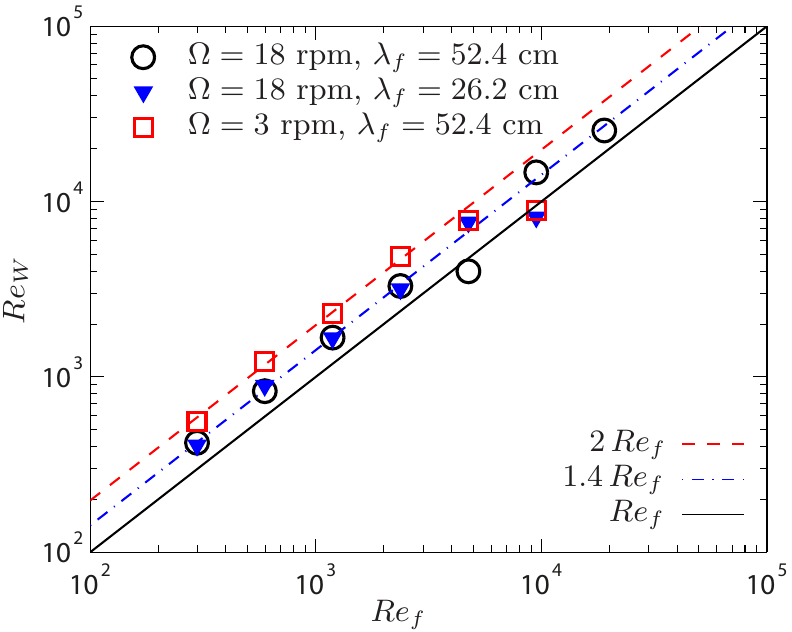}}
    \caption{Attractor Reynolds number $Re_W=W\lambda^2/\nu$ averaged
    over the first branch of the attractor (the one following the
    focusing reflection) as a function of the forcing Reynolds number
    $Re_f=U_f\lambda_f/\nu$. Square symbols correspond to experiments
    at ($\Omega=3$~rpm,~$\lambda_f=52.4$~cm), triangles to
    ($\Omega=18$~rpm,~$\lambda_f=26.4$~cm), and circles to
    ($\Omega=18$~rpm,~$\lambda_f=52.4$~cm).}\label{fig:ReW}
\end{figure}

Figure~\ref{fig:lambdaA} altogether allows us to state that the
attractor wavelength $\lambda$ and vorticity amplitude $W$ follow
the scaling laws predicted by the linear model but modified in the
non-linear regime by prefactors function of the forcing Reynolds
number
\begin{eqnarray}\label{eq:scalinglambdaW1}
\lambda&=&\ell^{2/3}L_0^{1/3}\,f(Re_f),\\
W&=&\frac{U_f\lambda_f}{\ell^{4/3}L_0^{2/3}}\,g(Re_f).\label{eq:scalinglambdaW2}
\end{eqnarray}
When the wavelength $\lambda$ predicted
by~(\ref{eq:scalinglambdaW1}) is larger than $\lambda_f/\gamma$, no attractor
can develop: a cutoff is
therefore expected
in~(\ref{eq:scalinglambdaW1}-\ref{eq:scalinglambdaW2}) when
$Re_f\geq f^{-1}(\lambda_f/\gamma\ell^{2/3}L_0^{1/3})$. As we will
see in the following, the thickening of the attractor and the
decrease of its relative amplitude when $Re_f$ increases above
$Re_f\simeq 1\,000$ is correlated to the onset of a triadic
resonance instability of the attractor. This instability drains
energy from the mode at $\sigma_0$ toward lower frequency modes.
For the mode at $\sigma_0$, the instability can be seen as an
additional dissipation to the viscous dissipation. A rudimentary
but simple way to account for this additional dissipation is to
replace the fluid viscosity by a turbulent viscosity $\nu_t
\propto U_f\lambda_f$. Doing so in the viscous length $\ell\sim
\nu^{1/2}$ appearing in
Eqs.~(\ref{eq:scalinglambdaW1}-\ref{eq:scalinglambdaW2}) leads to
$f(Re)= Re_f^{1/3}$ and $g(Re)= Re_f^{-2/3}$. Reporting the
laws~(\ref{eq:scalinglambdaW1}-\ref{eq:scalinglambdaW2}) with
these expressions in Fig.~\ref{fig:lambdaA}(a-b) provides an
excellent description of the attractor wavelength and amplitude,
confirming the relevance of the concept of turbulent viscosity to
understand the non-linear wave attractor. We note that no
numerical prefactor have been used when reporting
Eqs.~(\ref{eq:scalinglambdaW1}-\ref{eq:scalinglambdaW2}) in
Fig.~\ref{fig:lambdaA}.

In Fig.~\ref{fig:ReW}, we finally report the Reynolds number of
the attractor defined as $Re_W=W\lambda^2/\nu$ which is shown to
increase linearly with the forcing Reynolds number $Re_f$, over
the whole studied range. The ratio $Re_W/Re_f$ indeed seems to be
nearly constant for a given rotation rate: it is remarkably almost
unaffected by the onset of the attractor instability at
$Re_f\simeq 1\,000$. The ratio $Re_W/Re_f$ is nevertheless slowly
dependent on $\Omega$ with $Re_W/Re_f \simeq 2.0$ for
$\Omega=3$~rpm and $Re_W/Re_f \simeq 1.4 $ for $\Omega=18$~rpm.
Since one would expect $Re_W/Re_f \simeq 1$ if a simple and single
reflection of the forced wave is observed, the ratio $Re_W/Re_f$
can be seen as a quantifier of the presence of an attractor.
Following (\ref{eq:scalinglambdaW1}-\ref{eq:scalinglambdaW2}), one
has $Re_W/Re_f=g(Re_f)f(Re_f)^2$. The weak but clear dependence of
$Re_W/Re_f$ with the rotation rate $\Omega$ that we report here shows that
$f=\lambda/(\ell^{2/3}L_0^{1/3})$ and $g=W\ell^{4/3}L_0^{2/3}/(\lambda_f U_f)$
are weakly dependent on the cavity Ekman number $Ek=\nu/2\Omega H^2$ in
addition to the leading dependence on the Reynolds number $Re_f$. Since this
weak dependence does not involve the forcing wavelength $\lambda_f$ and
amplitude $U_f$, it might be related to the physics of Ekman viscous boundary
layers on the walls of the cavity. We are however currently not able to propose
an explanation for this behavior which is weak but significant.

\subsection{Triadic resonance instability}

\begin{figure}
    \centerline{\includegraphics[width=10cm]{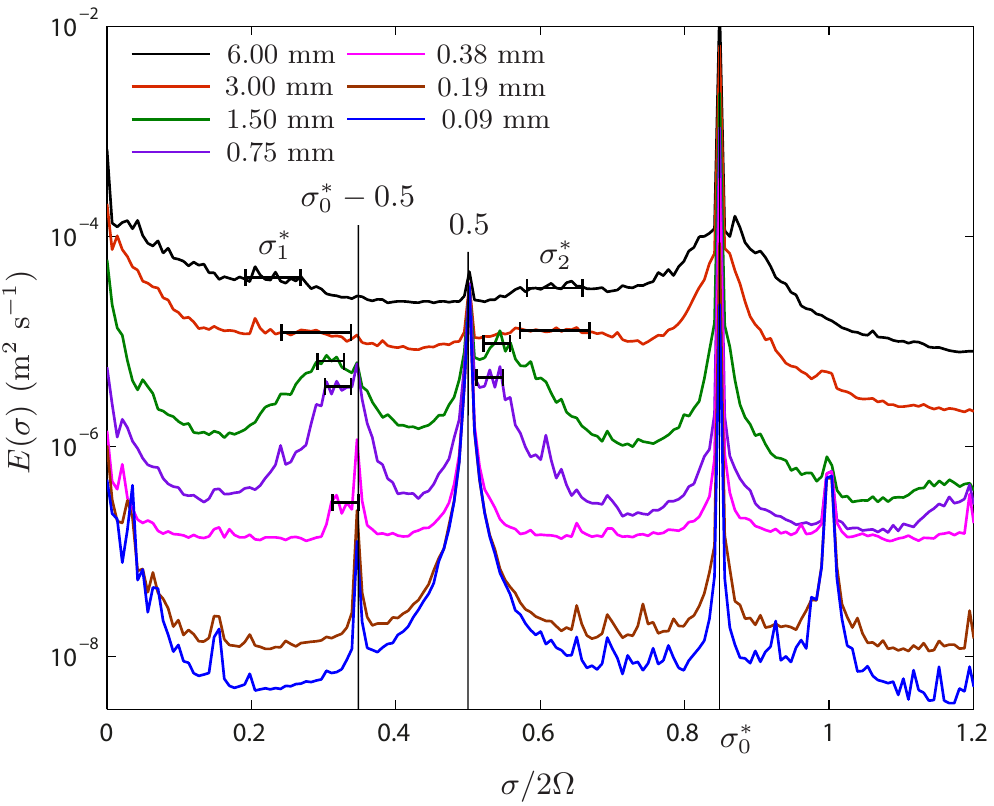}}
    \caption{Temporal power spectral density $E(\sigma)$
    (Eq.~\ref{eq:psd}) as a function of the normalized frequency
    $\sigma^*=\sigma/2\Omega$ for all forcing amplitudes $A$ at
    $\Omega=18$~rpm and $\lambda_f=52.4$~cm. For each spectrum, the
    horizontal error bars indicate the frequency intervals around
    which the subharmonic bumps are centered. These intervals
    correspond to the frequencies reported in Table~\ref{tab:freq}. We
    have highlighted with vertical lines three other energy peaks at
    frequencies $\sigma^*_0=\sigma_0/2\Omega$, $\sigma^*=0.5$ (i.e.
    $\sigma=\Omega$) and
    $\sigma^*=\sigma_0^*-0.5$.}\label{fig:spectrefreq}
\end{figure}

To further understand the non-linear evolution of the flow beyond
the instability threshold of the attractor, we report in
Fig.~\ref{fig:spectrefreq} the temporal energy spectrum
$E(\sigma)$ (Eq.~\ref{eq:psd}) for all experiments at
$\Omega=18$~rpm and $\lambda_f=52.4$~cm. Beyond $A=0.38$~mm
($Re_f\simeq 1\,200$) at which the linear prediction for the
attractor thickness and amplitude start to fail, we observe the
emergence of two subharmonic bumps in the spectrum. The
frequencies $\sigma_1$ and $\sigma_2$ around which the bumps are
centered are consistent with a triadic resonance with the forcing
frequency $\sigma_0$, i.e. $\sigma_1+\sigma_2=\sigma_0$, as can be
seen in the Table~\ref{tab:freq}. We recall that the energetic
peak at frequency $\sigma^*=0.5$, i.e. $\sigma=\Omega$, observed
for the lower amplitudes $A$ corresponds to the ``tilt-over'' flow
forced by the Earth rotation which induces a Coriolis force on the
fluid moving in the laboratory~\cite{Boisson2012,Triana2012}. This
peak probably hides the expected second energy bump in the
experiments at $A=0.38$~mm for which we report only one
subharmonic frequency $\sigma_1$. We also highlight that the sharp
peak observed at $\sigma^*=0.35$, i.e. $\sigma=\sigma_0-\Omega$,
for the low forcing experiments corresponds to the interaction of
this ``tilt-over'' flow with the forcing frequency $\sigma_0$.

\begin{table}
    \begin{tabular}{p{2cm} p{1.1cm} p{1.1cm} p{1.7cm} p{1.7cm}
        p{1.7cm} p{1.7cm} p{1.7cm}}
    \hline \hline
    $A$~(mm) & 0.09 & 0.19 & 0.38 & 0.75 & 1.50 & 3.00 & 6.00\\
    \hline $Re_f$ & 300 & 600 & 1\,190 & 2\,380 & 4\,750 & 9\,500 &
    19\,000 \\
    \hline $\sigma_1^*$ & --- & --- & 0.32$\pm$0.01 &
    0.32$\pm$0.02 & 0.31$\pm$0.02 & 0.29$\pm$0.05 & 0.23$\pm$0.04 \\
    \hline $\sigma_2^*$ & --- & --- & --- & 0.53$\pm$0.02 &
    0.54$\pm$0.02 & 0.62$\pm$0.05& 0.64$\pm$0.04\\
    \hline $(\sigma_1+\sigma_2)/\sigma_0$ & --- & --- & --- & $1.00 \pm 0.05$ &
    $1.00 \pm 0.05$ & $1.07 \pm 0.12$ & $1.02 \pm 0.09$ \\
    \hline \hline
    \end{tabular}
    \caption{Normalized center frequencies
    $\sigma_1^*=\sigma_1/2\Omega$ and $\sigma_2^*=\sigma_2/2\Omega$ of
    the subharmonic bumps observed in the temporal energy spectra for
    $\Omega=18$~rpm and $\lambda_f=52.4$~cm
    (Fig.~\ref{fig:spectrefreq}) as a function of the forcing
    amplitude $A$. Empty cells (i.e. with ``---'') correspond to cases
    when no bump is observable.}\label{tab:freq}
\end{table}

In any case, we can highlight that the emergence of the
subharmonic instability through a triadic resonance illustrated by
Fig.~\ref{fig:spectrefreq} is fully correlated with the increase
of the attractor lengthscale and to the damping of its normalized
amplitude revealed in Figs.~\ref{fig:alllaw} and
\ref{fig:lambdaA}. The subharmonic bumps in the temporal spectra are wide, a
feature that
was already reported for the triadic resonance instability of an
experimental plane inertial wave in~\cite{Bordes2012}. It confirms
that there is a specificity for experimental inertial waves with
respect to internal waves~\cite{Bourget2013,Scolan2013} and
numerical inertial waves~\cite{Jouve2014} for which triadic
instability produces two precise frequencies.

In Fig.~\ref{fig:spectrefreq}, for the two largest forcing
amplitudes $A=3$~mm and $6$~mm ($Re_f \simeq 9\,500$ and
$19\,000$), the subharmonic bumps become hardly distinguishable.
We believe that this last feature does not mean that the
instability has vanished since the total energy stored in the
modes at subharmonic frequencies is still significant but spread
over large frequency ranges. The horizontal error bars shown in
Fig.~\ref{fig:spectrefreq} aim at representing qualitatively the
uncertainty on the determination of the central frequency of the
subharmonic bumps $\sigma_1^*$ and $\sigma_2^*$. Thus, considering
with precaution the spectra for the two largest amplitudes in
Fig.~\ref{fig:spectrefreq}, we can note that the separation
between the bumps center frequencies $\sigma_1^*$ and $\sigma_2^*$
seem to increase with $A$, $\sigma_1^*$ and $\sigma_2^*$ going
further away from $\sigma_0^*/2$. The bumps seem at the same time
to get wider whereas their amplitudes $E(\sigma_{1,2})$
progressively decrease with $A$ relatively to the base level of
the spectrum. These results are in discrepancy with the temporal
spectra reported in~\cite{Jouve2014} for numerical simulations of
inertial wave attractor in which $\sigma_1^*$ and $\sigma_2^*$ are
clearly defined by sharp peaks and tend toward $\sigma_0^*/2$ as
the forcing amplitude increases. These points remain to be
understood and will be discussed in the section~\ref{sec:discut}.

To confirm that the flow components associated to the bumps at
$\sigma_1$ and $\sigma_2$ are composed of inertial waves, we
search in the following for the spatio-temporal signature of their
dispersion relation. To do so, we compute for each experiment the
normalized spatio-temporal power spectral density of the velocity
field as
\begin{eqnarray}\label{eq:psd2}
E'(k_x,k_z,\sigma)=\frac{|\tilde{u}'_j(k_x,k_z,\sigma)|^2}{\langle|\tilde{u}'_j(k_x,k_z,\sigma)|^2\rangle},
\end{eqnarray}
where
\begin{eqnarray}
\tilde{u}'_j(k_x,k_z,\sigma)=\frac{1}{(2\pi)^3}\int_0^T\int_0^{L_x}\int_0^{H}
u_j(x,y_0,z,t)e^{-i(\sigma t + k_x x + ik_z z)} \,dt dx dz
\end{eqnarray}
is the spatio-temporal Fourier transform of $u_j(x,y_0,z,t)$ with
$j=x,z$ and the angular brackets represent the average over
wavenumber space (normalization by the energy at $\sigma$).

\begin{figure}
     \centerline{\includegraphics[width=0.95\textwidth]{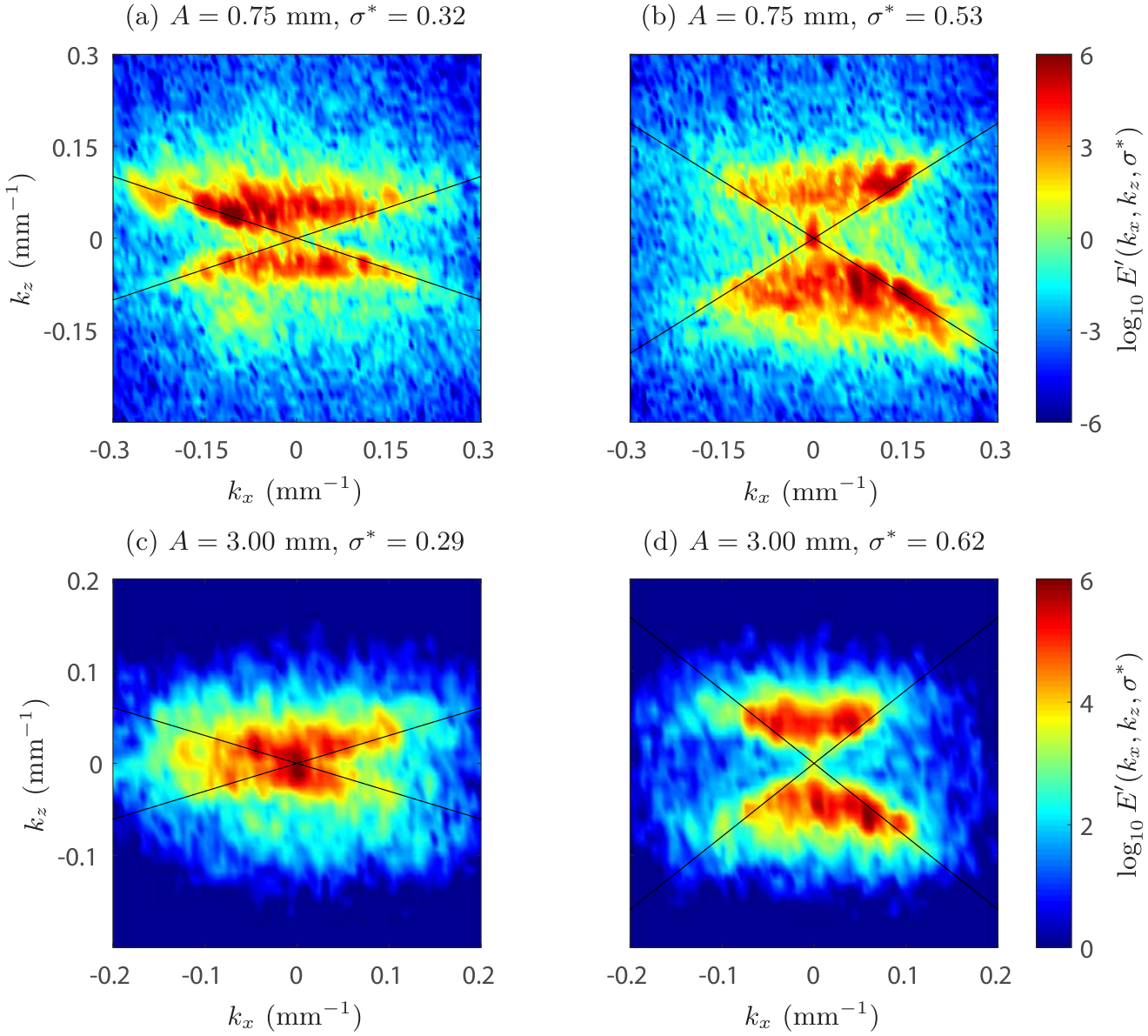}}
\caption{Spatio-temporal power spectral density
$E'(k_x,k_z,\sigma)$ (Eq.~\ref{eq:psd2}) for the experiments at
$\Omega=18$~rpm, $\lambda_f=52.4$~cm and $A=0.75$~mm (a-b,
$Re_f\simeq 2\,400$) or $A=3.00$~mm (c-d, $Re_f\simeq
9\,500$). The selected frequencies for each experiment correspond
to the center frequencies of the subharmonic bumps in their
temporal spectrum, i.e. $\sigma^*=0.32$ (a) and $\sigma^*=0.53$
(b) for $A=0.75$~mm and $\sigma^*=0.29$ (c) and $\sigma^*=0.62$
(d) for $A=3.00$~mm (see Table~\ref{tab:freq}). In each panel,
black lines correspond to the dispersion relation
$|\sigma^*|=|k_z|/(k_x^2+k_z^2)^{1/2}$ of inertial waves with
$k_y=0$.}\label{fig:Spatiotemp0_750mm}
\end{figure}

In Fig.~\ref{fig:Spatiotemp0_750mm}, we report this
spatio-temporal spectrum $E'(k_x,k_z,\sigma)$ for the experiments
at $\Omega=18$~rpm, $\lambda_f=52.4$~cm and $A=0.75$~mm
($Re_f\simeq 2\,400$) or $A=3.00$~mm ($Re_f\simeq 9\,500$), for
the respective center frequencies of the bumps in their temporal
spectrum. Black lines represent the dispersion relation
$|\sigma^*|=|k_z|/(k_x^2+k_z^2)^{1/2}$ of 2D inertial waves
invariant in the direction~$y$ i.e. with $k_y=0$.  In such a
representation, inertial waves with their wavevector in the
measurement plane will appear through energy concentration on the
black lines.

\begin{figure}
    \centerline{\includegraphics[width=0.95\textwidth]{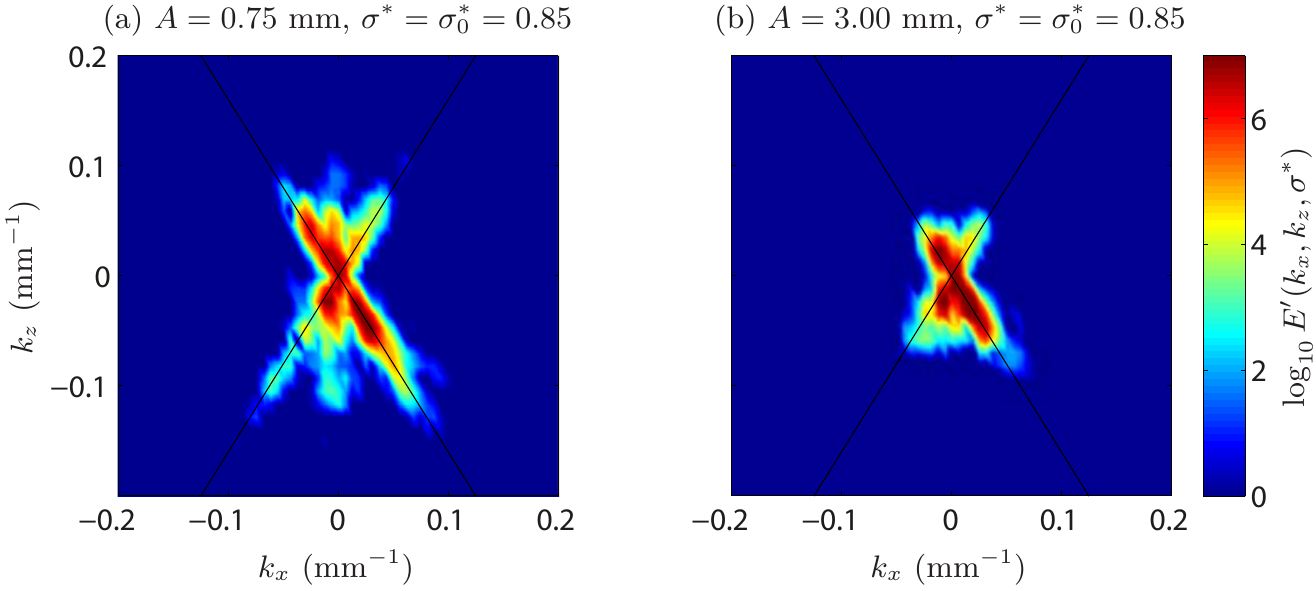}}
     \caption{Spatio-temporal power spectral density $E'(k_x,k_z,\sigma)$
(Eq.~\ref{eq:psd2}) at the forcing frequency $\sigma=\sigma_0$ for
the experiments at $\Omega=18$~rpm, $\lambda_f=52.4$~cm and
$A=0.75$~mm (a, $Re_f\simeq 2\,400$) and $A=3.00$~mm (b,
$Re_f\simeq 9\,500$). In each panel, black lines correspond to the
dispersion relation $|\sigma^*|=|k_z|/(k_x^2+k_z^2)^{1/2}$ of
inertial waves with $k_y=0$.}\label{fig:Spatiotempbis}
\end{figure}

In Fig.~\ref{fig:Spatiotemp0_750mm}, the energetic regions of the
spatio-temporal spectra ressemble a sandglass with several maxima
of energy along the black lines. This shows that the flow
component at frequencies $\sigma_1$ and $\sigma_2$ is composed of
inertial waves of which a significant proportion has a non-zero
wavevector component along $y$ direction. Indeed, the general
expression of the dispersion relation is
$|\sigma^*|=|k_z|/(k_x^2+k_y^2+k_z^2)^{1/2}$, and inertial waves
with $k_y \neq 0$ will show up with energy in the regions between
the two black lines ($|\sigma^*|=|k_z|/(k_x^2+k_z^2)^{1/2}$) and
containing the axis $k_x=0$. Considering other forcing amplitude
$A$ and other frequencies~$\sigma^*$ inside the subharmonic bumps
of the temporal spectra leads to similar spatio-temporal spectra.
The only exception is the spatio-temporal spectra at the
forcing frequency $\sigma_0$ in which the energy is clearly
concentrated on the black lines as can be seen in
Fig.~\ref{fig:Spatiotempbis} (same experiments as in
Fig.~\ref{fig:Spatiotemp0_750mm}). This result confirms that the
attractor remains nearly two-dimensional, with $k_y\simeq 0$, in
the measurement plane $y=L_y/3$. For other frequencies
$\sigma^*$, the presence of waves with wavevector in the
measurement plane $y=y_0$ is natural since the triadic instability
of an attractor invariant in the $y$ direction leads to such waves
if no spontaneous breaking of symmetry appears. Waves found here
with $k_y \neq 0$ might be the fruit of the non-perfect
$y$-invariance of the experimental attractor due to the finite
size in the $y$ direction of the wavemaker and/or to the presence
of viscous boundary layers on the vertical walls of the cavity at
$y=0$ and $y=L_y$ which both should lead to some
three-dimensionality of the flow. This three-dimensionality might
also be at the origin of the previously highlighted differences
(bumps spreading in frequency and moving away from each other with
increasing $A$) of the temporal spectra with the simulations of
Jouve and Ogilvie~\cite{Jouve2014} which were strictly 2D,
invariant along $y$ direction.

\section{Conclusion}\label{sec:discut}

In this article, we have reported PIV measurements of the flow
generated by a large-scale harmonic inertial forcing in a
trapezoidal cavity with a tilted wall submitted to a global
rotation. In the linear regime, we observe a concentration of the energy along
a limit cycle inside the cavity as expected from the theory of internal wave
attractors. Our data shows that the
model, initially proposed by Rieutord~\textit{et~al.}~\cite{Rieutord2001}
followed by Grisouard~\textit{et~al.}~\cite{Grisouard2008} and Jouve and
Ogilvie~\cite{Jouve2014},
describing attractors as a portion of a self-similar wave beam
emitted by a virtual point source upstream of the tilted wall
accounts correctly for the measured values of the wavelength in
the attractor as well as for the scaling laws of the spatial decay
of its amplitude.

We have further explored the non-linear regime of the attractor.
The observed scenario is the following when increasing the forcing
amplitude. The attractor becomes unstable beyond a forcing
Reynolds number of $Re_f \simeq 1\,200$. This instability feeds
inertial waves gathered around two subharmonic
frequencies $\sigma_1$ and $\sigma_2$ resonant with the attractor
frequency $\sigma_0$. This triadic resonance instability is
accompanied by a thickening in size and a damping in relative
amplitude of the attractor as the forcing amplitude grows above
the instability threshold. In parallel, the two bumps
corresponding to the subharmonic waves in the temporal spectrum
have their central frequencies $\sigma_1$ and $\sigma_2$ gradually
moving away from $\sigma_0/2$ while the bumps spread over wider
ranges of frequencies tending to build a continuum of energy in
frequency.

In~\cite{Brouzet2017}, from similar experiments with internal gravity
waves, Brouzet~\textit{et al.} also report an increase of the
attractor wavelength and a reduction of its relative amplitude
when the attractor becomes unstable via a triadic resonance. They
interpreted their results by introducing a turbulent viscosity
accounting for the fact the instability creates a sink of energy
for the attractor. In this article, by considering data for two
different rotation rates $\Omega$ and for two forcing wavelengths
$\lambda_f$, we have demonstrated that the attractor mean
wavelength $\lambda$ and vorticity amplitude $W$ follow scaling
laws predicted by the linear attractor model even in the
non-linear regime if one uses the turbulent viscosity
$\nu_t=U_f\lambda_f$ based on the forcing velocity $U_f$ and
wavelength $\lambda_f$ in place of the fluid kinetic viscosity.
This framework eventually predicts that the attractor wavelength
$\lambda$ and vorticity $W$ follow power laws $\lambda \sim
Re_f^{1/3}$ and $W\sim Re_f^{-2/3}$ with the forcing Reynolds
number $Re_f$ beyond the onset of the triadic instability which
scalings are in clear agreement with our data.

Regarding the subharmonic waves produced by the instability, it is
worth highlighting two major differences with previous
results~\cite{Jouve2014,Scolan2013,Bourget2013,Brouzet2017}. The
first one relies on the fact that the instability provides energy
to large ranges of frequencies and not to two precise subharmonic
frequencies. This weak selectivity is not specific to attractors
since it has also been reported for a plane inertial wave
beam~\cite{Bordes2012}. It is in clear discrepancy with the strong
selectivity reported for numerical inertial
waves~\cite{Jouve2014}. It might be related to the
three-dimensional nature of the velocity oscillations in inertial
waves: in a $y$-invariant ($k_y=0$) wave at frequency $\sigma$,
fluid particles describe circular translations in planes tilted by
an angle $\theta=\cos^{-1}(\sigma/2\Omega)$ and velocity
oscillations are observed in the three directions $x$, $y$ and
$z$. The vertical walls of the cavity are therefore incompatible
with such $y$-invariance in inertial wave experiments. This
feature is on the contrary absent in 2D numerical simulations of
inertial waves in which periodic boundary conditions are used in
the $y$ direction~\cite{Jouve2014}. In the case of internal
gravity waves~\cite{Scolan2013,Bourget2013,Brouzet2017}, for which
fluids particles oscillate only in the vertical plane ($x,z$), the
physical vertical walls $y=0$ and $y=L_y$ are compatible with
inviscid boundary conditions. The three-components character of
inertial waves interacting with physical walls therefore appears
as a good candidate to explain the weak selectivity of the triadic
instability. This could be tested by comparing the results of
numerical simulations with $y$ periodic boundary conditions and
with walls at $y=0$ and $y=L_y$. In this context, it is worth
noting that Manders and Maas~\cite{Manders2004} have studied the
three-dimensional structure of an experimental inertial wave
attractor forced by a libration perturbation of the global
rotation in a trapezoidal cavity. The dependence of the
experimental attractor in the $y$-direction is however probably
strongly different compared to the one in the data we report here
since the libration forcing is anti-symmetric for mirror
reflection with respect to the plane $y=L/2$, leading to a phase
shift of $\pi$ in the attractor between half-space $y<L/2$ and
half-space $y>L/2$. The libration forcing actually induces an
oscillating horizontal circulation in the plane $(x,y)$ with
strong horizontal velocities close to the vertical walls which is
also a feature absent in our experiments.

A last major difference with previous works is that the
instability of the inertial wave attractor reported here leads to
subharmonic waves with frequencies more and more remote from
$\sigma_0/2$ as the forcing amplitude grows. This behavior is the
opposite of the one reported in numerical simulations of an
inertial wave attractor in a tilted square~\cite{Jouve2014}. It is also in
apparent contradiction with
the theory of triadic instability of plane inertial/internal
waves~\cite{Koudella2006,Bordes2012} which predicts that the
frequencies of the two subharmonic waves tend toward $\sigma_0/2$
as the Reynolds number of the primary wave increases. This behavior is probably
the most intriguing that has been reported here.

\acknowledgments We acknowledge M. Rabaud and F. Moisy for
fruitful discussions, and J. Amarni, A. Aubertin, L. Auffray and
R. Pidoux for experimental help. This work has been supported by
the Agence Nationale de la Recherche through Grant ``DisET''
No.~ANR-17-CE30-0003.


\begin{thebibliography}{99}
\bibitem{phaseaverage}
The phase average of the time series $f(t)$ at frequency
$\sigma_0$ is defined by $\hat{f}(\varphi)=\sum_{n=1}^N
f((\varphi+n 2\pi)/\sigma_0)/N=\sum_{n=1}^N f(t+n2\pi/\sigma_0)/N$
where $N$ is the total number of period $2\pi/\sigma_0$ of the
signal and $\varphi=\sigma_0 t$.
\end{thebibliography}
\end{document}